\documentclass[journal=cmatex,manuscript=article]{achemso}
\usepackage[utf8]{inputenc}
\usepackage{amsmath}

\usepackage{physics}
\usepackage{makecell}
\usepackage{mathtools}
\usepackage{amsfonts}
\usepackage{amssymb}
\usepackage{graphicx}
\usepackage{wrapfig}
\usepackage{textcomp}
\usepackage{color}
\usepackage[dvipsnames]{xcolor}
\usepackage{multirow}
\usepackage{comment}
\usepackage{gensymb}
\usepackage{chemformula}

\title{Chirality-tunable non-linear Hall effect}

\author{Nesta Benno Joseph}
\affiliation{Solid State and Structural Chemistry Unit, Indian Institute of Science, Bangalore 560012, India}
\author{Arka Bandyopadhyay}
\affiliation{Solid State and Structural Chemistry Unit, Indian Institute of Science, Bangalore 560012, India}
\author{Awadhesh Narayan}
\email{awadhesh@iisc.ac.in}
\affiliation{Solid State and Structural Chemistry Unit, Indian Institute of Science, Bangalore 560012, India}

\begin{document}

\begin{tocentry}
\begin{center}
\includegraphics[width=7 cm]{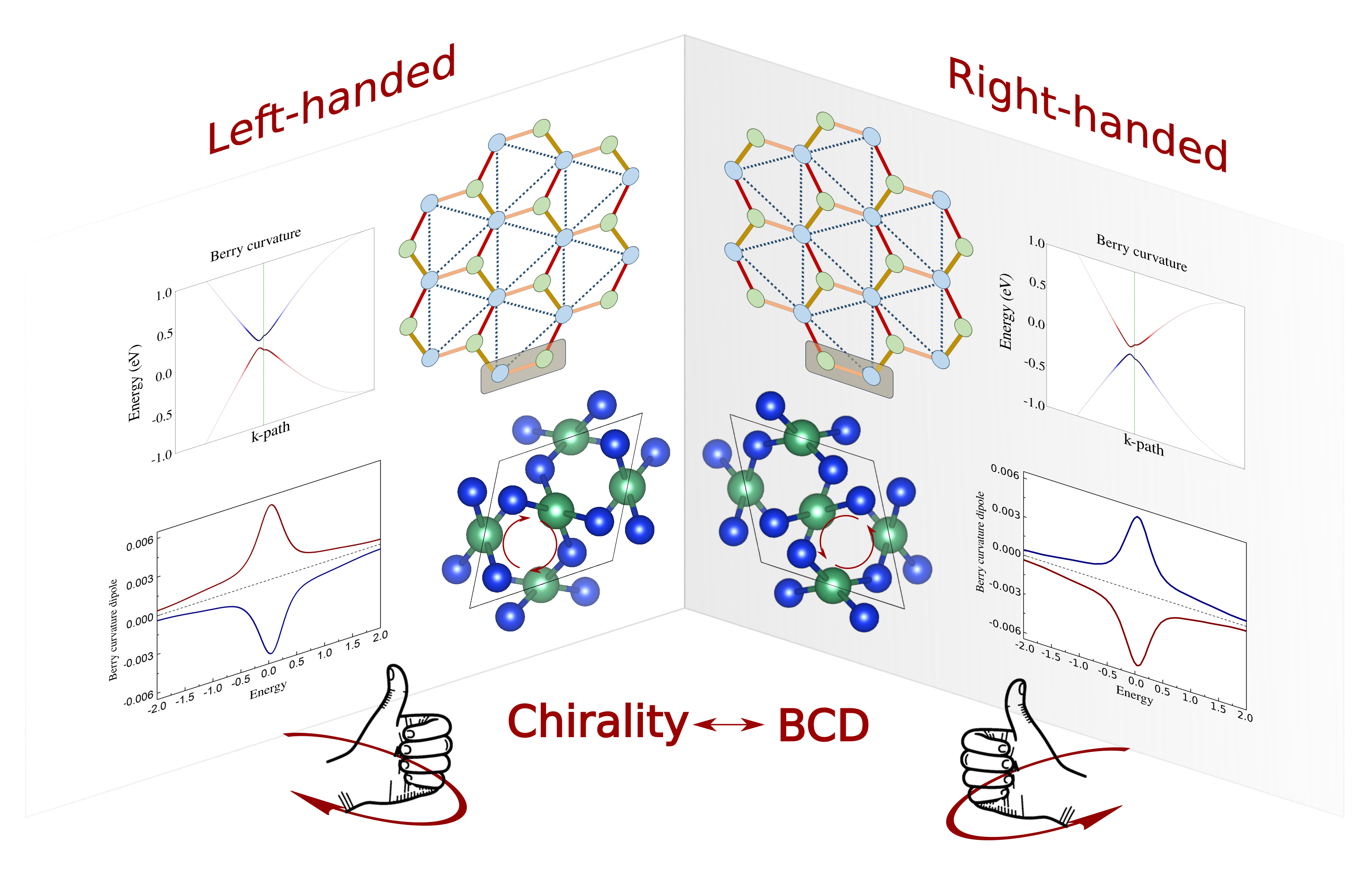}
\end{center}

\end{tocentry}

%\date{\today}

\begin{abstract}
The non-linear generalization of the Hall effect has recently gained much attention, with a rapidly growing list of non-centrosymmetric materials that display higher-order Hall responses under time-reversal invariant conditions. 
The intrinsic second-order Hall response arises due to the first-order moment of Berry curvature -- termed Berry curvature dipole -- which requires broken inversion and low crystal symmetries. 
Chiral materials are characterized by their lack of improper symmetries such as inversion, mirror plane, and roto-inversion.
Owing to this absence of symmetries, in this work, we propose chiral systems as ideal platforms to study the Berry curvature dipole-induced non-linear Hall effects.
We use state-of-the-art first-principles computations, in conjunction with symmetry analyses, to explore a variety of chiral material classes  -- metallic \ch{NbSi2}, semiconducting elemental Te, insulating HgS, and topological multifold semimetal CoSi.
We present the emergence and tunability of the Berry curvature dipole in these chiral materials. 
In particular, we demonstrate that the two enantiomeric pairs exhibit an exactly opposite sign of the Berry curvature dipole.
We complement our \textit{ab initio} findings with a general tight-binding minimal model and give estimates for non-linear Hall voltages, which are experimentally accessible.
Our predictions put forward chiral materials as an emerging class of materials to realize non-linear Hall phenomena and highlight an as-yet-unexplored aspect of these systems.
\end{abstract}

\maketitle

\clearpage 

\section{\label{sec:intro} Introduction}

Chirality is a well-established property of asymmetry or handedness observed in certain structures or systems~\cite{kelvin1894molecular,wagniere2007chirality}. 
The absence of mirror planes, center of inversion and rotation-reflection axes are characteristics of a chiral object.
The lack of these symmetry elements leads to non-superimposable mirror images of the system, much like how the left hand cannot be perfectly overlaid onto the right hand. 
A chiral system is also often associated with handedness (left-handed/right-handed) and helicity (clockwise/anti-clockwise), with mirror images having opposite handedness/helicity.
The concept of chirality is widespread across various disciplines, including chemistry, biology, and physics~\cite{wagniere2007chirality,barron2008chirality}. 
In molecular systems, enantiomers are non-superimposable mirror images of stereoisomers with similar physical and chemical properties except for the direction in which they rotate plane-polarized light.
In materials, chirality can manifest in molecular arrangements, crystal structures, or electronic configurations, leading to unique properties and behaviors. 
A crystal structure is said to be chiral if its space group consists of only proper operations -- inversion, mirror plane, roto-inversion and glide are all improper operations.
Hence, the only allowed symmetry elements are rotations and translations. 
Most chiral crystals crystallize in space groups that are enantiomeric pairs of each other.
Just as the enantiomers of a chiral molecule are related by a mirror reflection, enantiomorphic pairs of a chiral crystal are related by screw rotations.
Chiral materials often exhibit distinct optical, electronic, or mechanical characteristics, making chirality an important consideration in the design and study of materials with applications ranging from pharmaceuticals to advanced materials~\cite{blaser2013chirality,brandt2017added,dor2013chiral,naaman2015spintronics,liang2022enhancement}. 
In recent years, a number of intriguing aspects of chiral materials have been discovered, including interplay with topology~\cite{chang2018topological,hasan2021weyl,yang2023monopole}, phonons~\cite{zhang2015chiral,zhu2018observation,ishito2023truly,ueda2023chiral,romao2023phonon,lange2023negative}, and ferroelectricity~\cite{hu2020chiral,fu2022multiaxial,das2024family}.

The celebrated Hall effect -- the development of a voltage transverse to the direction of current flow as well as the applied magnetic field -- has led to a new paradigm in condensed matter physics and materials science~\cite{chien2013hall}. 
It has revealed deep connections to topology and geometry~\cite{cage2012quantum,maciejko2011quantum,chang2023colloquium}. In recent years, the notion of Hall effect has been extended to time-reversal symmetric systems. 
Such Hall signals are symmetry allowed at non-linear orders and require breaking of inversion symmetry~\cite{du2021nonlinear,ortix2021nonlinear,bandyopadhyay2024non}. Sodemann and Fu connected the second-order Hall response to the first moment of the Berry curvature and termed it the Berry curvature dipole (BCD)~\cite{sodemann2015quantum}. 
Since their initial proposal, a number of materials have been predicted to exhibit a BCD and the resulting non-linear Hall effects.
These include two-dimensional transition metal dichalcogenides~\cite{son2019strain,you2018berry,xiao2020two,zhou2020highly,joseph2021tunable,zhang2018electrically,joseph2021topological,he2021giant,jin2021strain}, topological semimetals~\cite{zhang2018berry, zeng2021nonlinear,singh2020engineering,pang2024tuning,chen2019strain}, graphene-analogs~\cite{bandyopadhyay2022electrically,bandyopadhyay2023berry}, and organic compounds~\cite{kiswandhi2021observation}, to highlight a few. 
Remarkable progress has also been made in realizing these effects experimentally.
Non-linear Hall signals have been recently revealed in transport measurements on van der Waals materials~\cite{ma2019observation,kang2019nonlinear,ma2022growth,huang2023giant,kang2023switchable,ho2021hall,huang2023intrinsic,he2022graphene,sinha2022berry,zhong2023effective}, topological crystalline insulators~\cite{zhang2022pst,nishijima2023ferroic}, and Dirac and Weyl semimetals~\cite{shvetsov2019nonlinear,kumar2021room}.
Furthermore, the non-linear Hall effect is expected to be of direct application in terahertz sensors~\cite{zhang2021terahertz} and non-linear devices~\cite{chien2013hall,ramsden2011hall}.

In this contribution, we propose chiral materials -- with their inherent broken inversion symmetry -- as intriguing platforms to study the BCD, and the resulting non-linear Hall effect.

As we discussed earlier, breaking of inversion symmetry is essential for obtaining non-zero BCD under time-reversal symmetric conditions. 
Furthermore, the crystal symmetry significantly influences the number of nonzero components. 
Therefore, given their property of universal absence of inversion symmetry and the lack of additional restrictions imposed by mirror planes, chiral crystals act as promising candidates for the generation of BCD and non-linear Hall currents. 
Our work sheds light on the unexplored interplay between the handedness of a chiral system and the sign of BCD components through the tight-binding method and first-principles calculations on real material systems. 
Starting with a general tight-binding model, we highlight the nature of the Berry curvature and its dipole in the enantiomers. 
We demonstrate that the two enantiomers of a chiral system show opposite signs of BCD and, therefore, opposite non-linear Hall effect signals. 
We then use state-of-the-art first-principles calculations and symmetry analyses to investigate four diverse chiral materials classes -- metallic \ch{NbSi2}, semiconducting elemental Te, insulating HgS, and topological multifold semimetal CoSi -- and establish the relation between the Berry curvature and its dipole in the enantiomorphs. 
Our findings suggest that BCD-induced non-linear Hall measurements offer a promising avenue for characterizing chiral materials and distinguishing between enantiomeric pairs.
Our work could motivate experimental investigations of chiral materials, in the near future, as suitable candidates for exploring non-linear Hall effects.

\section{\label{sec:methods} Computational methods}

Our first-principles calculations were carried out based on the density functional theory (DFT) framework as implemented in the {\sc quantum espresso} code~\cite{QE-2017,QE-2009}. 
Three chiral material candidates that crystalize in two enantiomeric space groups were chosen for our study -- trigonal Te, trigonal HgS (both having space groups $P3_121$ (152) and $P3_221$ (154)), and \ch{NbSi2} (space groups $P6_222$ (180) and $P6_422$ (181)).
A fourth candidate chosen is the topological multifold semimetal CoSi, crystalizing in the cubic space group $P2_13$ (198).
A kinetic energy cut-off of $40$, $40$, $50$, and $50$ Ry were considered for Te, HgS, \ch{NbSi2}, and CoSi, respectively, using the ultrasoft pseudopotentials~\cite{PhysRevB.41.7892} to describe the core electrons and including spin-orbit interactions in case of Te and CoSi. We used the Perdew-Burke-Ernzerhof (PBE) form for the exchange-correlation functional~\cite{perdew1996generalized}.
For trigonal Te, the use of PBE severely underestimates the bulk band gap. Since the nature of the Berry curvature, and hence the BCD, around Fermi level depends on the band gap of the system, we use the Heyd–Scuseria–Ernzerhof (HSE) screened hybrid functional~\cite{heyd2003hybrid} to accurately describe the band gap.
A semi-empirical Grimme's DFT-D2~\cite{DFT-D2} correction was included to account for the van der Waals interactions for both Te and HgS.
The Brillouin zone was sampled over a uniform $\Gamma$-centered $k$-mesh of $4\times4\times2$ (Te and HgS), $4\times4\times3$ (\ch{NbSi2}), and $4\times4\times4$ (CoSi).

To calculate Berry curvature and related properties, a Wannier-based tight-binding model was derived from the \textit{ab initio} calculations, using the {\sc wannier90} code~\cite{mostofi2014updated}. 
The Wannier functions (WFs) associated with a unit cell are given by~\cite{marzari1997maximally}

\begin{equation}
    \big|\textbf{R}_n\big \rangle = \frac{V}{(2\pi)^3} \int d\textbf{k}~e^{-i\textbf{k}\cdot\textbf{R}}~\big|\psi_{n\textbf{k}}\big \rangle,
\end{equation}

where $\big|\psi_{n\textbf{k}}\big \rangle$ is the Bloch state associated with band $n$ at momentum vector $\textbf{k}$. 
The maximally localized Wannier functions (MLWFs) are constructed by minimizing the total spread function, $\Delta$, of $J$ Wannier functions $|\textbf{R}_n \rangle$ and reads~\cite{marzari1997maximally}

\begin{equation}
    \Delta = \sum_{n=1}^J [\langle r^2 \rangle_n - \Bar{\textbf{r}}_n^2],
\end{equation}

where $\Bar{\textbf{r}}_n = \langle \textbf{0}_n|\textbf{r}|\textbf{0}_n \rangle$ and $\langle r^2 \rangle_n = \langle \textbf{0}_n|r^2|\textbf{0}_n \rangle$.
This procedure utilizing MLWFs does not always guarantee the correct description of centers and point symmetries of orbitals considered. 
A more optimal description and localization of orbitals may be obtained by using the selectively localized Wannier functions (SLWFs) approach~\cite{wang2014selectively}, where a subset of the Wannier functions $J' \le J$ with specified centers and point symmetries are generated. The modified spread function, with these constraints imposed on it, is given by 

\begin{equation}
    \Delta_c = \sum_{n=1}^{J'} [\langle r^2 \rangle_n - \Bar{\textbf{r}}_n^2 + \lambda_c(\Bar{\textbf{r}}_n - \textbf{r}_{0n})^2].
\end{equation}

Here the Lagrange multiplier, $\lambda_c$, is introduced to constrain the $n$-th Wannier center at $\Bar{\textbf{r}}_n$ to the desired center $\textbf{r}_{0n}$. This modified spread function is now minimized to obtain the SLWFs with greater localization of the selected subset of Wannier functions around a fixed center.  
We used the SLWFs approach in the case of trigonal Te to construct Wannier functions for the calculation of Berry curvature and BCD. 
The Wannier functions associated with \ch{NbSi2}, HgS, and CoSi were generated using the MLWFs approach.

From the constructed WFs, the BCD can be evaluated and has the form, 

\begin{equation}
    D_{ab} = \sum_{n}\int_{k}~f_n^0(\textbf{k})~\frac{\partial\Omega^n_b}{\partial k_a}d\textbf{k},
    \label{eq:bcdexpression}
\end{equation}

with $f_n^0$(\textbf{k}) being the equilibrium Fermi-Dirac distribution and $\Omega$, the Berry curvature given by

\begin{equation}\label{eqn:berry_curv}
    \Omega_a^n (\textbf{k}) =  2i\hbar^2 \sum_{m \neq n} \frac{\langle n|\Hat{v}_b|m\rangle \langle n|\Hat{v}_c|m\rangle}{(\epsilon_n - \epsilon_m)^2}.
\end{equation}

Here $\epsilon_m$ and $|m\rangle$ are the $m$-th energy eigenvalue and eigenvector of the Hamiltonian, $a,b,c\in\{x,y,z\}$ denote the direction, $\hat{v}_b$ and $\hat{v}_c$ are the velocity operators, and $\hbar$ is the reduced Planck's constant.
Crystal point symmetries impose constraints on the BCD tensor of the form

\begin{equation}\label{symmetry}
    D = \mathrm{det}(S)SDS^T,
\end{equation}

where $S$ is the orthogonal matrix that describes the point symmetry and $\mathrm{det}$ denotes the determinant. The non-zero components of BCD can be identified by analyzing this condition (Eq.~\ref{symmetry}).
We used our Wannier-based tight-binding Hamiltonian to compute the Berry curvature and associated BCD using the {\sc wannier-berri} code~\cite{tsirkin2021high}. 
The BCD calculations of the candidate materials were done at 40 K for \ch{NbSi2}, Te and HgS, and at 100 K for CoSi.
The calculation of Weyl chirality in Te system was carried out using WannierTools code~\cite{WU2017}.

%----------------------------------------Results and Discussions-----------------------------------------------------------

\section{Results and discussion}

\subsection{Chiral tight-binding model}

\begin{figure*}
%    \centering
    \includegraphics[scale=0.71]{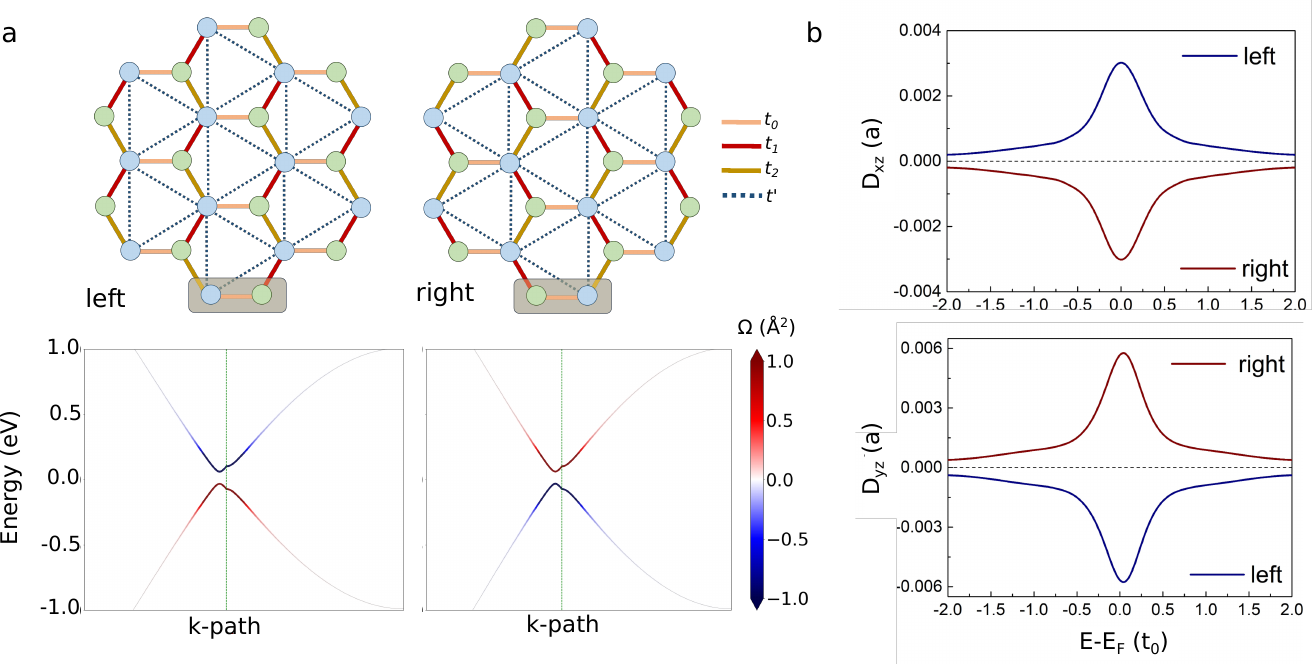}
    \caption{\textbf{Chiral tight-binding model}. (a) Left- and right-handed enantiomers (top) with the hopping terms indicated. The two systems are non-superimposable mirror images of each other. Two distinct sublattices A (blue) and B (green) of a unit cells are shaded in gray. The Berry curvature superimposed on the band structure (bottom). Note the reversal of the sign of the Berry curvature for the two structures. (b) The two non-zero components of BCD -- $D_{xz}$ and $D_{yz}$ -- of left- and right-handed systems. The sign of the BCD components are opposite for the two members of the enantiomer pair. The representative parameters used here are $t_0 = 1$, $t_1 = 1.01 \:t_{0}$, $t_2 = 0.99 \:t_0$, and $t^{\prime} = -0.01 \: t_0$.}
    \label{figure:TB model}
\end{figure*}

As we noted above, chiral crystals manifest a distinctive geometric property whereby their mirror image is non-superimposable onto the original crystalline structure. We note that both the original and mirror-image structures inherently lack inversion symmetry. This absence of an inversion center essentially contributes to the unique, non-identical nature of the mirror images, emphasizing the structural chirality inherent in these crystalline arrangements. To understand the non-linear Hall response of such structures, we introduce a minimal tight-binding Hamiltonian that captures all the essential features of a model chiral system [Fig.~\ref{figure:TB model} (a)]. The Hamiltonian, defined on a honeycomb lattice, reads     

\begin{equation}
    H = \sum_{\langle i,j\rangle} [t_{ij} \: a_{i}^{\dagger} b_{j} + h.c.] + \sum_{\langle\langle i,j\rangle\rangle} [g_{ij}^{l} \: a_{i}^{\dagger} a_{j} + g_{ij}^{r} \: b_{i}^{\dagger} b_{j} + h.c.].
    \label{eq:tightham}
\end{equation}

Here, the operators $a_{i}^{\dagger} \:(a_{i})$ and $b_{i}^{\dagger} \:(b_{i})$ denote the creation (annihilation) operators responsible for creating (annihilating) an electron at $i$-th site on sublattices A and B, respectively. Here $h.c.$ denotes the Hermitian conjugate. The symbols $\langle...\rangle$ and $\langle\langle...\rangle\rangle$ indicate the hopping interactions between nearest and next-nearest neighbors, similar to the hopping mechanisms observed in graphene. It is noteworthy, however, that in contrast to graphene, three adjacent bond lengths and their corresponding hopping parameters are distinct ($t_0$, $t_1$, $t_2$) without breaking the lattice periodicity. In addition, the next-nearest-neighbor hopping parameters between A and B sublattices are expressed as $g_{ij}^{l}$ and $g_{ij}^{r}$, respectively. We note that the strengths of the next nearest neighbor hopping integrals are equivalent, i.e., $g_{ij}^{l} = g_{ij}^{r} = t^{\prime}$. In the context of chirality, the hopping parameter $g_{i,j}^{r}$ ($g_{i,j}^{l}$) is deactivated for the left-handed (right-handed) enantiomer to preserve its distinctive chirality characteristics. 

The right and left enantiomers, shown in Fig.~\ref{figure:TB model} (a), are non-superimposable images of each other with broken inversion symmetry. Therefore, the electronic band structure of these systems exhibits a non-vanishing Berry curvature (only along the $z$ direction, perpendicular to the plane) even under time-reversal symmetric conditions. However, the sign of the Berry curvature of a particular band undergoes a sign flipping while going from a left-handed to a right-handed system, as presented in Fig.~\ref{figure:TB model} (a). The above analysis helps us to comprehend that the crystallographic chiral symmetry influences the sign of the Berry curvature of a particular band. Next, we delve into examining the resulting influence of the Berry curvature sign change on the direction of the BCD in a chiral lattice. The necessary symmetry-breaking conditions of the systems are achieved by means of the unequal values of the three neighboring hopping parameters and the next-nearest neighbor hopping exclusively originating from one sublattice. The expression given in Equation~\ref{eq:bcdexpression} reveals that two components of BCD -- $D_{xz}$ and $D_{yz}$ -- will exhibit non-zero contributions. Fig.~\ref{figure:TB model} (b) shows the non-zero BCD components as a function of the Fermi energy for the two enantiomers. We find that both the components of BCD undergo a sign reversal while switching from the left to the right enantiomer. These observations in our minimal model set the stage for exploring the effect of chirality on non-linear Hall response in our chosen classes of real chiral materials. 

\subsection{\ch{NbSi2}}

\begin{figure}
    \centering
    \includegraphics[scale=0.45]{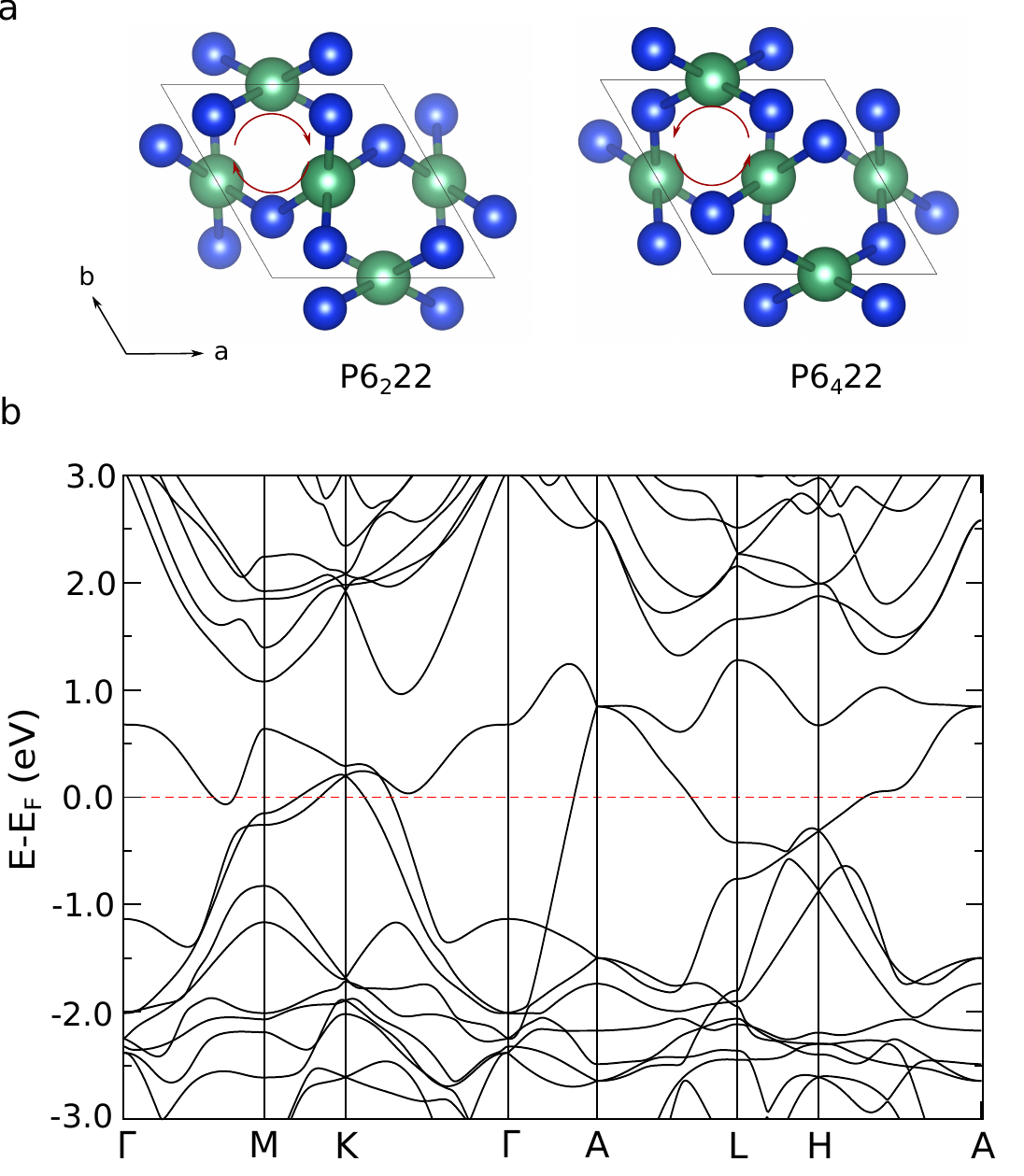}
    \caption{\textbf{Structure of \ch{NbSi2}}. (a) \ch{NbSi2} crystallizes into two enantiomeric pairs of space group $P6_222$ and $P6_422$. The clockwise and anti-clockwise orientations of Nb-Si-Nb chains are highlighted by the red arrows. The green and blue spheres represent Nb and Si atoms, respectively. (b) The electronic band structure of \ch{NbSi2} shows the metallic nature of the material. The two enantiomers exhibit an identical band diagram.}
    \label{fig:nbsi2_bands}
\end{figure}

\begin{figure*}
    \centering
    \includegraphics[scale=0.75]{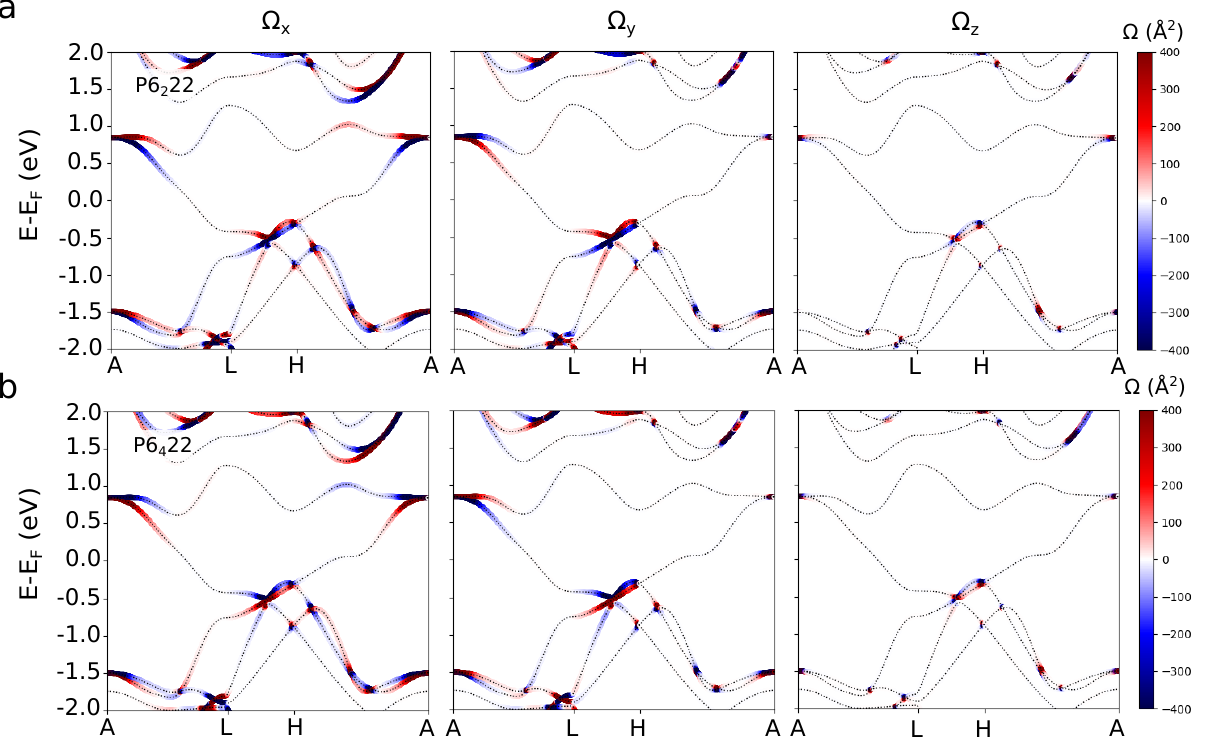}
    \caption{\textbf{Berry curvature associated with \ch{NbSi2} enantiomers}. The $x$ (left), $y$ (middle), and $z$ (right) components of the momentum-resolved Berry curvature, $\Omega$, of the two enantiomers (a) $P6_222$ and (b) $P6_422$ superimposed on the band structure along the $k_z=0.5\pi$ plane. The two pairs exhibit opposite signs of Berry curvature for all the components. Berry curvature hot spots are abundant and arise in the vicinity of band touching points.}
    \label{fig:nbsi2_bc}
\end{figure*}

\subsubsection{Structure and symmetry}

The first candidate material that we propose to study the chirality-induced BCD is \ch{NbSi2}, a member of the family of transition metal silicides (\ch{TSi2}) that crystallize in C40 structure. 
One way to examine the chirality in this material is the opposite helicity of the two pairs of Nb-Si-Nb chains intertwined along the $c$-axis, crystallizing in two enantiomorphic space groups, $P6_222$ (180) and $P6_422$ (181), as shown in Fig.~\ref{fig:nbsi2_bands} (a). 
The symmetries present in the system include a two-fold proper rotation ($C_2$) along [100], [010] and [001] axes, a six-fold screw rotation ($6_2$/$6_4$ rotation depending on the space group) along [001], and two-fold screw rotations along $a-b$ face diagonal (a screw rotation of $N_n$ corresponds to a rotation by $2\pi/N$ followed by a translation $\tau=n/N$ of the lattice point, with $\tau$ in fractional coordinates. $N_n$ translates to $N_m$ via inversion, with $m+n = N$). 
Imposing these symmetry constraints on the BCD components (Equation ~\ref{symmetry}), we discover that the off-diagonal elements go to zero. 
Furthermore, the diagonal elements are not independent, and the BCD tensor has the form

\begin{equation}\label{eqn:symmetry_NbSi}
    D =  \pm D_{xx}
\begin{bmatrix}
1 & 0 & 0 \\
0 & 1 & 0 \\
0 & 0 & -2
\end{bmatrix},
\end{equation}

with the sign of $D_{xx}$ determined by the chirality (left-handed versus right-handed) of the structure. We next investigate the electronic structure, Berry curvature, and BCD properties of the \ch{NbSi2} enantiomers.

\begin{figure}
    \centering
    \includegraphics[scale=0.6]{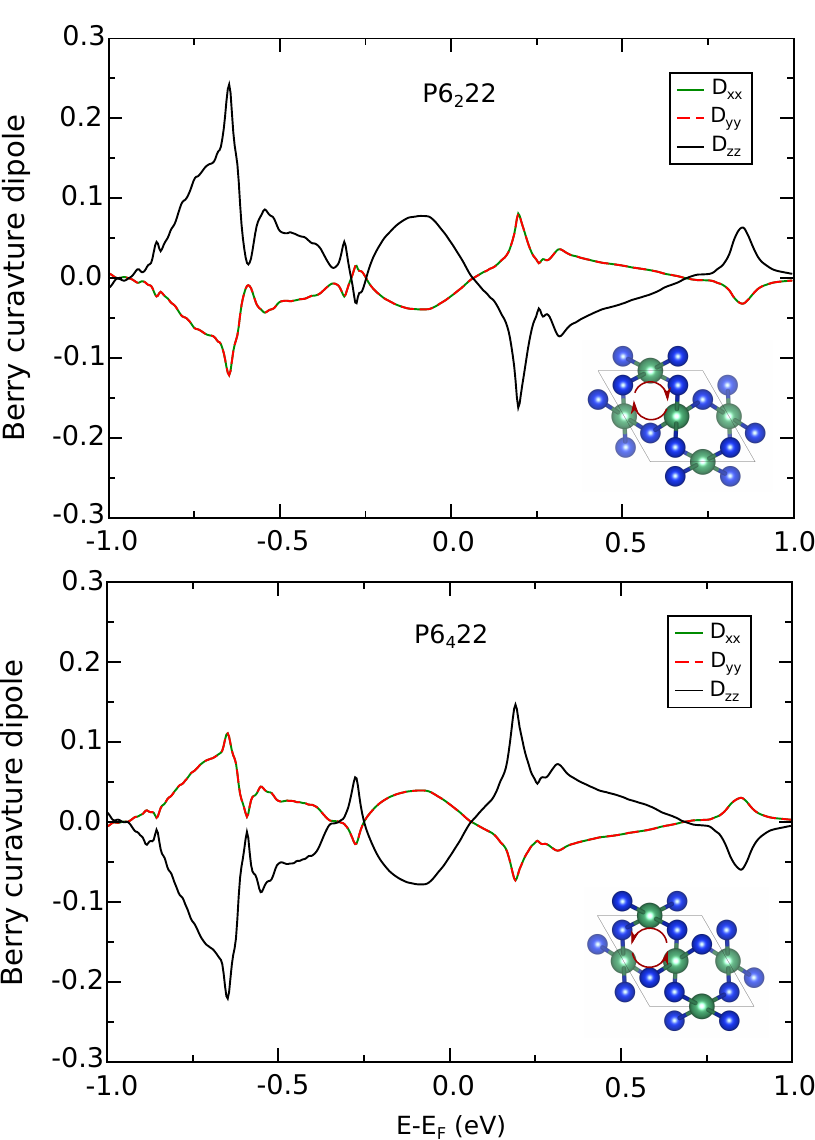}
    \caption{\textbf{Berry curvature dipole of \ch{NbSi2} enantiomers.}. The non-zero components of BCD -- $D_{xx}$, $D_{yy}$ and $D_{zz}$ -- associated with the pair of enantiomers of \ch{NbSi2}, plotted as a function of the Fermi level for $P6_222$ (top) and $P6_422$ (bottom). Only the diagonal components are non-zero due to the symmetries present and are related as $-D_{zz} = D_{xx} + D_{yy}$ since BCD is a traceless tensor. The enantionmeric pair displays opposite BCD and hence, opposite direction of non-linear Hall current under an applied electric field. The insets show the helical structures. Note that BCD is dimensionless in three dimensions.}
    \label{fig:nbsi2_bcd}
\end{figure}

\subsubsection{Electronic structure and Berry curvature properties}

The electronic band structure of \ch{NbSi2}, as shown in Fig.~\ref{fig:nbsi2_bands} (b), reveals that the system is metallic in nature. 
There exists a small electron pocket along $\Gamma$-M direction and larger hole pockets along M-K-$\Gamma$ and $\Gamma$-A-L directions. 
The states near the Fermi level are dominated by Nb $d$ orbitals, with minor contributions from Si $p$ orbitals.

We next computed the Berry curvature of \ch{NbSi2}. The momentum-resolved Berry curvature superimposed on the band structure is shown in Fig.~\ref{fig:nbsi2_bc}. 
We find that all three components -- $\Omega_x$, $\Omega_y$, and $\Omega_z$ -- show a number of hot spots in the vicinity of band crossing and anticrossing points.
The value of the Berry curvature exceeds 400 \AA{}$^2$, with the maximum value concentrated around high symmetry A point.
Noticeably, the $z$ component of the Berry curvature is less pronounced than the other two components.
This is the direction of the helices.
Strikingly, we find that all three of the Berry curvature components have the exactly opposite sign for the two enantiomers (see Fig.~\ref{fig:nbsi2_bc}). 

By using the symmetry analysis, we found that the off-diagonal elements of the BCD go to zero (Equation~\ref{eqn:symmetry_NbSi}).
Further, we found that the sum of two of the diagonal elements should be equal and opposite to the third component, since BCD is a traceless tensor. 
Hence, the diagonal elements of the BCD tensor in the case of \ch{NbSi2} follow the relations, $-D_{zz} = D_{xx} + D_{yy}$ and $D_{xx} = D_{yy}$. 
We present the three non-zero components of the BCD, as a function of the Fermi energy, for the two enantiomers in Fig.~\ref{fig:nbsi2_bcd}.
Several features are noteworthy.
Comparing the two enantiomers, we discover that all the components of the BCD undergo a sign reversal for the two.
This is in agreement with our proposed minimal tight-binding model and shows that reversal of BCD between enantiomers is a general feature of chiral systems.
Prominent peaks in the BCD are associated with a rapid variation in the Berry curvature.
This is particularly highlighted near $E-E_F=-0.6$ eV, where peaks are seen for all components of the BCD.
This occurs due to the rapidly varying Berry curvature in this energy range, as seen in Fig.~\ref{fig:nbsi2_bc}. 
The calculated absolute values of BCD at the Fermi level are $\sim|0.022|$ and $\sim|0.05|$ for $D_{xx}$ ($D_{yy}$) and $D_{zz}$ components, respectively. 
This value is comparable to previous reports on BCD for systems such as \ch{LiOsO3}~\cite{xiao2020electrical}, while being slightly smaller than the values computed for Weyl semimetals~\cite{zhang2018berry,kang2019nonlinear}.

\subsection{Trigonal Tellurium and HgS}

\begin{figure*}
    \centering
    \includegraphics[scale=0.56]{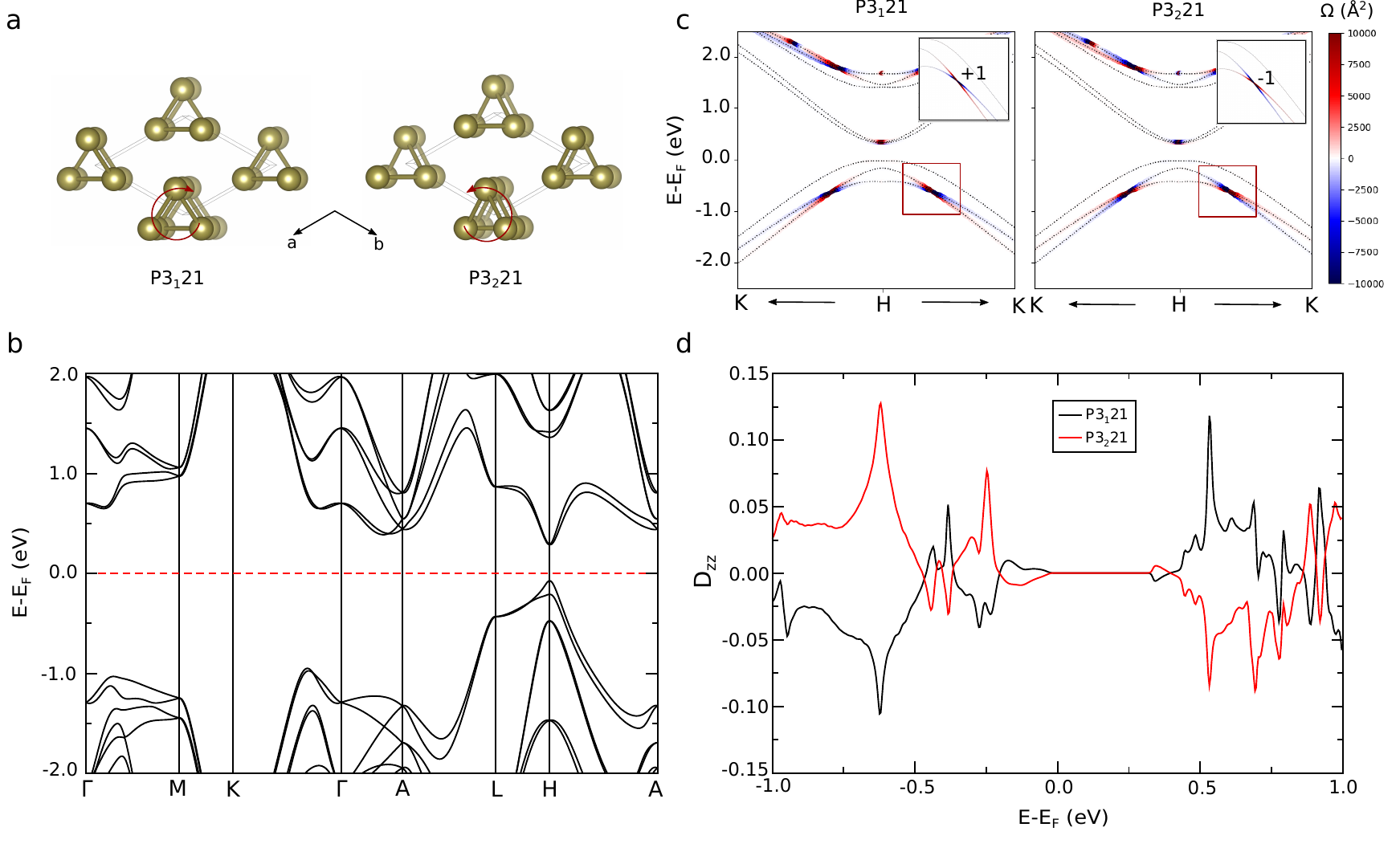}
    \caption{\textbf{Trigonal elemental Te}. (a) The structure of the two enantiomers of Te and their respective space groups. The red arrows represent the direction of helicity (clockwise/anti-clockwise) of Te helices. (b) The electronic band structure of Te. Elemental Te is a semiconductor with a bandgap of 0.36 eV. There also exist Weyl points around the H point near Fermi level. (c) The $z$ component of the momentum-resolved Berry curvature, $\Omega_z$, of the two enantiomers along the K$-$H$-$K direction of the Brillouin zone. The Berry curvature is largely concentrated near the Weyl points and shows opposite signs for the enantiomeric pairs. Insets reveal the Weyl points along K$-$H having opposite signs of both Berry curvature and calculated Weyl chirality ($\pm 1$). (d) The $D_{zz}$ component of BCD as a function of Fermi energy. The BCD flips sign depending on the chirality of the structure. Other components of BCD also show a similar enantiomer-dependent nature.}
    \label{fig:Te}
\end{figure*}

\subsubsection{Structure and symmetry}

\begin{figure*}
    \centering
    \includegraphics[scale=0.7]{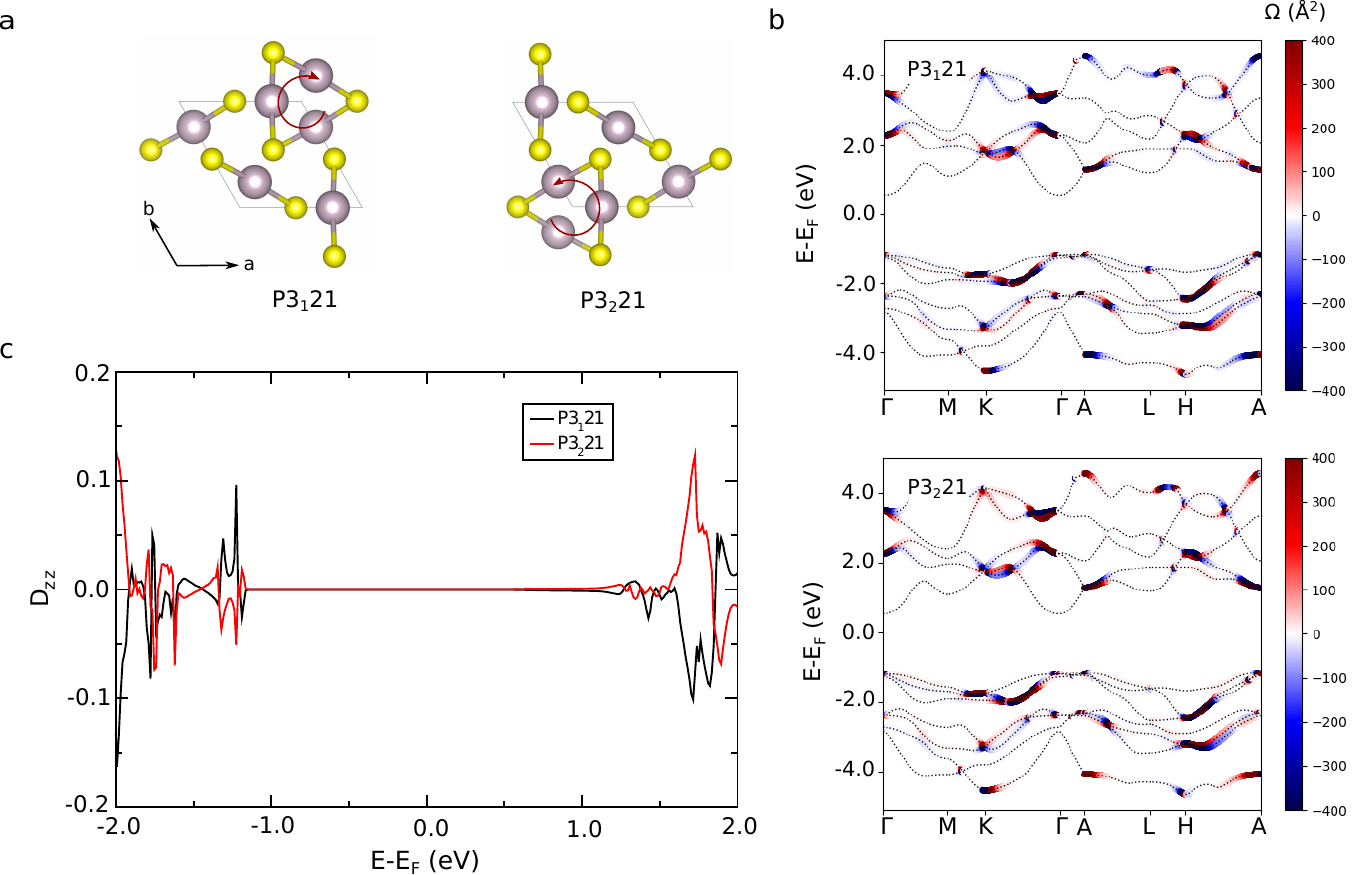}
    \caption{\textbf{Trigonal $\alpha$-HgS}. (a) The two enantiomers of $\alpha$-HgS with opposite chirality with the corresponding space groups. The violet and yellow spheres represent Hg and S atoms, respectively. (b) The electronic band structure and $x$-component of the momentum-resolved Berry curvature, $\Omega_x$, distribution for the two enantiomers. This phase of HgS is an insulator with a calculated bandgap of $\sim$1.7 eV. (c) The BCD as a function of the Fermi energy for the two enantiomeric systems. BCD exhibits an opposite sign depending on the chirality of the structure.}
    \label{fig:HgS}
\end{figure*}

Next, we turn our attention to trigonal elemental Te and insulating HgS. 
Both have a similar non-centrosymmetric crystal structure, with one-dimensional helical strands of Te (HgS) bound together by van der Waals forces, arranged in a trigonal lattice.  
The space group associated with the structures can be either $P3_121$ (152) or $P3_221$ (154), depending on the left-handed or right-handed screw axis of the helical chains, leading to chirality in these materials [see Fig.~\ref{fig:Te} (a) and Fig.~\ref{fig:HgS} (a)]. 
These pairs of enantiomers have a two-fold proper rotation axis ($C_2$) along the [110], a three-fold screw rotation ($3_1$/$3_2$ rotation for $P3_121$/$P3_221$) axis along [001] directions, and two-fold screw rotation axes along [100] and [010]. The BCD tensor for both systems takes the form,

\begin{equation}
    D =  \pm D_{xx}
\begin{bmatrix}
1 & 0 & 0 \\
0 & 1 & 0 \\
0 & 0 & -2 \\
\end{bmatrix}.
\label{eq.nine}
\end{equation}

The sign of $D_{xx}$ is determined by the chirality of the system, as we shall see next. 

\subsubsection{Electronic structure and Berry curvature properties}

We begin by examining the electronic structure and Berry curvature of Te enantiomers.
Te is a highly-studied system, owing to being an elemental chiral material, with a number of well-established chirality-dependent properties~\cite{tsirkin2018gyrotropic,sakano2020radial,liu2023electricalmagnetochiralanisotropytrigonal}.
It is a $p$-type narrow gap semiconductor at ambient conditions, with a bandgap of $\sim$0.32 eV~\cite{anzin1977measurement}. 
The presence of Weyl nodes around the Fermi level gives it a non-trivial topology, and this material is often categorized as a "Weyl semiconductor"~\cite{hirayama2015weyl,nakayama2017band}. 
Theoretical calculations using the PBE form of the exchange-correlation functional underestimate the bandgap of Te to 30 meV (compared to the experimental bandgap of 0.32 eV), and hence, in this study, we calculated the electronic structure and other properties using the HSE screened hybrid functional. 
Figs.~\ref{fig:Te} (b) and (c) show the band structure of Te and the Weyl points present around high symmetry H point in the valence and conduction bands.
The HSE band structure reveals a band gap of $\sim$0.36 eV, in good agreement with the experimental value.

Analyzing the Berry curvature associated with the two enantiomers of Te, we find that it is largely concentrated around the Weyl points at high symmetry H point, Weyl points along the K-H line, and other regions with bands coming close together [as expected from Equation~\ref{eqn:berry_curv}]. 
Fig.~\ref{fig:Te} (c) shows the $z$ component of the Berry curvature, $\Omega_z$, for both the enantiomers.
We discover that the sign of the Berry curvature is exactly opposite for the two enantiomers, clearly visible from the inset of Fig.~\ref{fig:Te} (c).
Remarkably, we also note that the calculation of Weyl chirality of the band-touching points yields opposite values -- if the Weyl chirality of one of the Weyl points is +1 in one enantiomer, the same Weyl point has a chirality of -1 in the other.
Our finding highlights an interesting connection between real space (structural) and momentum space (topological) chirality.

We present the $D_{zz}$ component of BCD as a function of the Fermi energy, for both $P3_121$ and $P3_221$ structures, in Fig.~\ref{fig:Te} (d).
From further calculations of the BCD, we find that the signs of all non-zero components of BCD flip from one enantiomer to the other, in agreement with the conclusions from our chiral tight-binding model. 
The large peak in the BCD around $E-E_{F}\approx-0.6$ eV is due to the large concentration of Berry curvature density near the Weyl point, highlighted in the inset of Fig.~\ref{fig:Te} (c).
Somewhat smaller, yet prominent peaks are also observed near  $E-E_{F}\approx-0.2$ eV.
Given that elemental Te naturally exists as a $p$-type material, we expect non-linear Hall signals to be readily measurable without the requirement of any additional doping.
Furthermore, the synthesis methods to grow pure Te enantiomers are by now mature, and thus our proposal of opposite signs of BCD in enantiomers could be experimentally tested in the near future.
Very recently, chirality-dependent non-linear conductivity has been measured in Te~\cite{niu2023tunable,suarez2024odd}.
However, our predicted BCD-driven non-linear Hall response still awaits experimental confirmation.

Structurally very similar to Te, the trigonal phase of HgS, known as $\alpha$-HgS, is an insulating material with an experimental bandgap of $\sim$2.1 eV (our calculated bandgap $\sim$1.7 eV). 
Fig.~\ref{fig:HgS} (b) shows the $x$ component of Berry curvature, $\Omega_x$, superimposed on the band structure of the two enantiomers. 
It can be seen that, similar to \ch{NbSi2} and trigonal Te discussed earlier, the sign of Berry curvature is opposite in case of the two enantiomers across the Brillouin zone.
The BCD component, $D_{zz}$, plotted as a function of Fermi energy, for both enantiomers is presented in Fig.~\ref{fig:HgS} (c).
The BCD has an oscillating nature, both in the valence band and conduction band manifolds, with the peaks/dips corresponding to Berry curvature hot spots.
The values are particularly pronounced near the valence band edge, which is suggestive of a larger non-linear Hall signal upon hole doping compared to electron doping.
As observed for our other proposed systems, we find that the BCD shows a reversal of signs for the two enantiomers. 
We note here that chiral phonons in $\alpha$-HgS were recently measured using circularly polarized Raman scattering~\cite{ishito2023truly}.
Large single crystals of $\alpha$-HgS can be grown, which makes us hopeful that our chirality-dependent non-linear Hall effect may be measurable in this system.
However, doping~\cite{masse1978study} will be required to tune the Fermi level out of the gap and allow transport measurements.
In this regard, it is worth mentioning that $n$-doped quantum dots of $\beta$-HgS have been successfully synthesized~\cite{jeong2014air}.

\subsection{Strain-induced BCD in chiral CoSi}

\begin{figure*}
    \centering
    \includegraphics[scale=0.59]{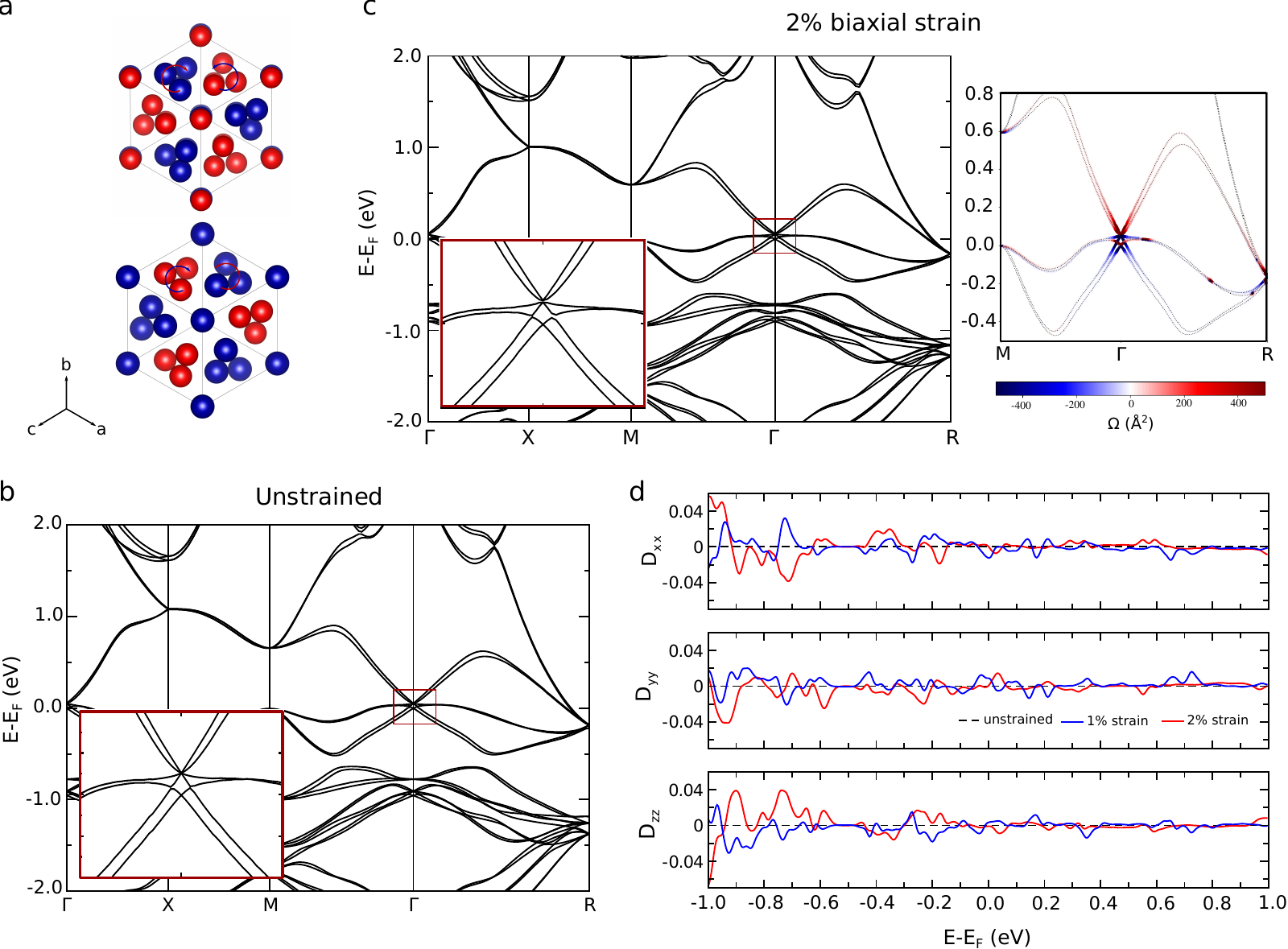}
    \caption{\textbf{Strain-induced BCD in CoSi}. (a) The two enantiomers of Cosi, both crystalizing in $P2_13$ space group, but with opposite chirality of the Co (blue) and Si (red) sublattices. (b) The electronic band structure of unstrained CoSi. Inset shows the four-fold degenerate bands present at $\Gamma$. (c) The electronic band structure and momentum-resolved Berry curvature distribution ($\Omega_z$) of CoSi under $2\%$ biaxial strain. Inset shows the lifting of degeneracy due to the broken $C_3$ rotation. The Berry curvature has a high concentration near $\Gamma$. (d) The $D_{xx}, D_{yy}$ and $D_{zz}$ components of BCD as a function of the Fermi energy for unstrained (black dashed), 1\% (blue solid) and 2\% (red solid) strained systems. Broken $C_3$ symmetry due to biaxial strain leads to non-zero BCD in this material.}\label{fig:CoSi}
\end{figure*}

\subsubsection{Structure and Symmetry}

CoSi crystallizes in space group $P2_13$ (198) and is among a large number of compounds having a chiral structure in a cubic space group. 
Unlike other candidate systems we discussed so far, these materials does not exist in an enantiomeric pair of space groups but rather crystallize as a racemate. 
When viewed along the [111] direction, the Co and Si sublattices form helices with opposite chirality [see Fig.~\ref{fig:CoSi} (a)]. 
Hence, the enantiomeric pair is usually denoted by RL and LR, RL meaning the Co sublattice is R type while Si sublattice is L type and vice versa for LR. 
Both RL and LR enantiomorphs crystalize with the same space group $P2_13$. 
In addition to $2_1$ screw rotation along [100], [010] and [001] axes, and three-fold screw rotations along $\langle111\rangle$ diagonals, CoSi has a three-fold proper rotation ($C_3$) along [111] direction.
The presence of this $C_3$ axis leads to the vanishing of all components of BCD in CoSi.
However, we propose that the application of a biaxial strain breaks this rotation symmetry and hence, we can obtain non-zero BCD.
This symmetry-breaking strategy is a general one to obtain finite BCD is all such compounds having a chiral structure in this cubic space group.
This is illustrated in the following section.

\subsubsection{Electronic structure and Berry curvature properties}

CoSi is a widely known topologically non-trivial material that hosts multifold fermions~\cite{tang2017multiple,rao2019observation}. 
Fig.~\ref{fig:CoSi} (b) shows the electronic band structure of CoSi. 
The four-fold degeneracy of bands at high symmetry $\Gamma$ point is protected by the three-fold rotation and two-fold screw rotation symmetries present in the system.
As we discussed, this same three-fold symmetry also leads to all components of BCD being identically zero in CoSi. 
%Application of a biaxial strain in the $a-b$ plane breaks this $C_3$ rotation~\cite{bose2021strain}. 
The application of a biaxial strain in the $ab$ plane breaks the cubic symmetry of CoSi, thereby breaking the three-fold rotation ($C_3$) present along the [111] direction~\cite{bose2021strain}.
As a result, the four-fold degeneracy of the band is lifted, as seen in Fig. 7 (c) for applied 2\% biaxial strain. 
We find that the Berry curvature is largely concentrated around the $\Gamma$ point due to the appearance of small band gaps and band anti-crossing features arising from the biaxial strain. 
From our symmetry analysis, we find that due to the broken $C_3$ rotation, the system is now allowed to have non-zero $D_{xx}, D_{yy}$, and $D_{zz}$ components of BCD (the other components are zero). 
For applied 1\% and 2\% biaxial strain, the variation of BCD components as a function of chemical potential is shown in Fig. 7 (d). 
All three non-zero components of BCD are highly oscillating, which is the direct consequence of a large number of band anticrossings and rapid variation of Berry curvature in the system.
Strain leads to multiple modifications in the band structure and the BCD is dependent on the details of band anticrossings, energy gaps between the bands, and the associated variation in Berry curvature. Therefore, a general trend in the dependence of BCD on strain values appears difficult to ascertain.
Overall, one can conclude that by means of a biaxial strain, chiral CoSi will display a strain-induced BCD and a resulting non-linear Hall response.
Finally, we note that the BCD values will have opposite signs for the RL and LR enantiomeric pairs, thereby resulting in opposite non-linear Hall signals.

\section{Estimation of non-linear Hall currents}

After presenting our candidate chiral materials, in this section, we estimate the strength of the non-linear Hall response that is expected in these systems. 
An oscillating external electric field $\vec{\mathcal{E}}(t) = \text{Re}[\vec{\mathcal{E}}_0 e^{i\omega t}]$ will effectively induce a second-order electric current, $j_{\alpha} = \text{Re}[J_{\alpha}^{(0)} + J_{\alpha}^{(2\omega)} e^{i2\omega t}]$, in systems characterized by non-vanishing BCD \cite{sodemann2015quantum}. Therefore, the response current can be decomposed into two distinct constituents -- a stationary component, denoted as the rectified current, given by $J_{\alpha}^{(0)} = \chi_{\alpha \beta \gamma} \vec{E}_{\beta} \vec{E}_{\gamma}^{*}$, and a dual-frequency oscillating component, termed the second harmonic, represented as $J_{\alpha}^{(2 \omega)} = \chi_{\alpha \beta \gamma} \vec{E}_{\beta} \vec{E}_{\gamma}$. 
Under time-reversal symmetric conditions, the nonlinear conductivity tensor, $\chi_{\alpha \beta \gamma}$, depends upon the momentum derivative of the Berry curvature across occupied states or the BCD, obtained as

\begin{equation}
    \chi_{\alpha \beta \gamma} =  \frac{\epsilon_{\alpha \gamma \delta} \: e^3 \tau}{2 \hbar^{2} (1+i\omega \tau)} D_{\beta \delta}.
    \label{eq:chi}
\end{equation}

Here the variables $e$, $\tau$, $\epsilon_{\alpha \gamma \delta}$, and $D_{\beta \delta}$ signify the electron charge, scattering time, Levi-Civita symbol, and BCD component (described in Equation~\ref{eq:bcdexpression}), respectively. 
Here, $\alpha, \beta, \gamma,\delta \in \{x,y,z\}$. 
In typical experimental arrangements, as illustrated in Refs.~\cite{ma2019observation,kang2019nonlinear}, the relaxation time, $\tau$, usually falls within the picosecond range, while the alternating current frequency is adjustable across the spectrum of 10 to 1000 Hz. 
Hence, we can neglect the frequency dependence in the denominator, as the product of $\omega$ and $\tau$ is significantly smaller than 1, $\omega \tau \ll 1$~\cite{roy2022non,saha2023nonlinear}. 
With this approximation, we can utilize the tensor representation of the BCD provided in Equations~\ref{eqn:symmetry_NbSi} and \ref{eq.nine} to obtain the following current density components

\begin{align}
    j_{x} & = \frac{3 e^3 \tau }{2 \hbar^2} (2 D_{xx}) \mathcal{E}_{z} \mathcal{E}_{y}, \nonumber \\
    j_{y} & = - \frac{3 e^3 \tau }{2 \hbar^2} (2 D_{xx}) \mathcal{E}_{z} \mathcal{E}_{x}, \nonumber \\
    j_{z} & = 0.
\label{eq:diffcontri1}
\end{align}

Let us consider a typical sample, with dimensions $10 \times 5 \times 3 \: (\mu \text{m})^3$, and isotropic resistivity along all directions (for simplicity), $\rho_x = \rho_y = \rho_z = 40 \: \mu \Omega \text{cm}$. 
These specific parameters yield the following values of current density, when the potential differences are chosen to be $V_x = V_y = V_z= 2$ V,

\begin{equation}
\begin{split}
    j_{x}  &= \frac{3 e^3 \tau }{2 \hbar^2} (2 D_{xx}) \mathcal{E}_{z} \mathcal{E}_{y} = \frac{3 e^3 \tau }{ \hbar^2} \frac{V_z V_y}{L_z L_y}(D_{xx}) \\ &= 29.83\times 10^{10} (D_{xx}) \mathrm{A/m^2},  \nonumber \\
    j_{y}  &= - \frac{3 e^3 \tau }{2 \hbar^2} (2 D_{xx}) \mathcal{E}_{z} \mathcal{E}_{x}  =  -\frac{3 e^3 \tau }{ \hbar^2} \frac{V_z V_x}{L_z L_x}(D_{xx}) \\ & = - 14.92\times 10^{10} (D_{xx})\mathrm{A/m^2} , \nonumber \\
    j_{z} & = 0.
\end{split}
\end{equation}
\label{eq:diffcontri2}

These current densities induce a voltage drop in the sample of the order of (1.19 $D_{xx}$) V and (0.30 $D_{xx})$ V along $x$ and $y$ directions, respectively. 
Plugging in a typical value of $D_{xx}=0.05$, calculated for our proposed chiral systems, the obtained values are well within the measurable range of current experimental capabilities.
This should allow for a precise examination of our proposal of the crystal chirality-dependent non-linear Hall signals. 
We note that the non-linear Hall current amplifies with higher BCD values, which may be obtained by suitably tuning the Fermi level, and alters its direction in accordance with the sign reversal of the BCD. 
Therefore, the non-linear Hall response could be used as a direct and useful tool for probing the handedness of an enantiomer.

\section{Summary and outlook}

In summary, we proposed chiral materials, with their intrinsic broken inversion symmetry, as promising platforms for exploring non-linear Hall responses. 
Using a general tight-binding model, we demonstrated that the two enantiomers of a chiral system show opposite signs of the Berry curvature as well as the BCD. 
We investigated three chiral materials -- \ch{NbSi2}, Te and HgS -- that crystallize in enantiomeric space groups as viable candidates for our proposal.
We identified the different symmetry elements present in these systems and noted, by symmetry analysis, that all three materials exhibit non-zero diagonal components of BCD.
Using state-of-the-art first-principles calculations, we showed that the change in chirality in these materials will lead to a reversal of sign in the BCD and, hence, the non-linear Hall signals. 
We note that this is true for all materials that crystallize in the 22 chiral (or 11 enantiomeric pairs of) space groups.
We presented a generally applicable symmetry-breaking strategy, via strain, to obtain finite BCD in chiral materials, where non-linear Hall response is symmetry-forbidden.
We exemplified our approach using the chiral topological material CoSi, by rotational symmetry breaking through a biaxial strain.
We note that the magnitude of BCD obtained for candidate materials are of the same order of magnitude as observed for other three-dimensional materials such as Weyl semimetals \ch{MoTe2}~\cite{singh2020engineering}, \ch{WTe2}~\cite{zhang2018berry} and TaAs~\cite{zhang2018berry}, polar metal \ch{LiOsO3}~\cite{xiao2020electrical}, and Rashba system BiTeI~\cite{facio2018strongly}.
An estimation of the non-linear Hall voltages suggests that direct measurements of our proposal should be possible with existing lock-in techniques.
We are hopeful that our work will motivate experimental and theoretical investigations on chiral materials as promising platforms for exploring non-linear Hall effects. 

\section*{Acknowledgements}

We thank S. Bhowal, A. Bose, M. Jain, H. R. Krishnamurthy, S. Mandal, A. Mahajan, N. Spaldin, and D. Varghese for valuable discussions and related collaborations. N.B.J. acknowledges Prime Minister’s Research Fellowship (PMRF) for support. A.B. acknowledges the financial support from Indian Institute of Science IoE postdoctoral fellowship. A. N. acknowledges support from DST CRG grant (CRG/2023/000114). 

\bibliography{references.bib}

\providecommand{\latin}[1]{#1}
\makeatletter
\providecommand{\doi}
  {\begingroup\let\do\@makeother\dospecials
  \catcode`\{=1 \catcode`\}=2 \doi@aux}
\providecommand{\doi@aux}[1]{\endgroup\texttt{#1}}
\makeatother
\providecommand*\mcitethebibliography{\thebibliography}
\csname @ifundefined\endcsname{endmcitethebibliography}  {\let\endmcitethebibliography\endthebibliography}{}
\begin{mcitethebibliography}{90}
\providecommand*\natexlab[1]{#1}
\providecommand*\mciteSetBstSublistMode[1]{}
\providecommand*\mciteSetBstMaxWidthForm[2]{}
\providecommand*\mciteBstWouldAddEndPuncttrue
  {\def\EndOfBibitem{\unskip.}}
\providecommand*\mciteBstWouldAddEndPunctfalse
  {\let\EndOfBibitem\relax}
\providecommand*\mciteSetBstMidEndSepPunct[3]{}
\providecommand*\mciteSetBstSublistLabelBeginEnd[3]{}
\providecommand*\EndOfBibitem{}
\mciteSetBstSublistMode{f}
\mciteSetBstMaxWidthForm{subitem}{(\alph{mcitesubitemcount})}
\mciteSetBstSublistLabelBeginEnd
  {\mcitemaxwidthsubitemform\space}
  {\relax}
  {\relax}

\bibitem[Kelvin(1894)]{kelvin1894molecular}
Kelvin,~W. T.~B. \emph{The molecular tactics of a crystal}; Clarendon Press, 1894; ISBN: 1358333793, 9781358333798\relax
\mciteBstWouldAddEndPuncttrue
\mciteSetBstMidEndSepPunct{\mcitedefaultmidpunct}
{\mcitedefaultendpunct}{\mcitedefaultseppunct}\relax
\EndOfBibitem
\bibitem[Wagni{\`e}re(2007)]{wagniere2007chirality}
Wagni{\`e}re,~G.~H. \emph{On chirality and the universal asymmetry: reflections on image and mirror image}; John Wiley \& Sons, 2007; ISBN: 9783906390383\relax
\mciteBstWouldAddEndPuncttrue
\mciteSetBstMidEndSepPunct{\mcitedefaultmidpunct}
{\mcitedefaultendpunct}{\mcitedefaultseppunct}\relax
\EndOfBibitem
\bibitem[Barron(2008)]{barron2008chirality}
Barron,~L.~D. Chirality and life. \emph{Strategies of Life Detection} \textbf{2008}, 187--201\relax
\mciteBstWouldAddEndPuncttrue
\mciteSetBstMidEndSepPunct{\mcitedefaultmidpunct}
{\mcitedefaultendpunct}{\mcitedefaultseppunct}\relax
\EndOfBibitem
\bibitem[Blaser(2013)]{blaser2013chirality}
Blaser,~H.-U. Chirality and its implications for the pharmaceutical industry. \emph{Rendiconti Lincei} \textbf{2013}, \emph{24}, 213--216\relax
\mciteBstWouldAddEndPuncttrue
\mciteSetBstMidEndSepPunct{\mcitedefaultmidpunct}
{\mcitedefaultendpunct}{\mcitedefaultseppunct}\relax
\EndOfBibitem
\bibitem[Brandt \latin{et~al.}(2017)Brandt, Salerno, and Fuchter]{brandt2017added}
Brandt,~J.~R.; Salerno,~F.; Fuchter,~M.~J. The added value of small-molecule chirality in technological applications. \emph{Nature Reviews Chemistry} \textbf{2017}, \emph{1}, 0045\relax
\mciteBstWouldAddEndPuncttrue
\mciteSetBstMidEndSepPunct{\mcitedefaultmidpunct}
{\mcitedefaultendpunct}{\mcitedefaultseppunct}\relax
\EndOfBibitem
\bibitem[Dor \latin{et~al.}(2013)Dor, Yochelis, Mathew, Naaman, and Paltiel]{dor2013chiral}
Dor,~O.~B.; Yochelis,~S.; Mathew,~S.~P.; Naaman,~R.; Paltiel,~Y. A chiral-based magnetic memory device without a permanent magnet. \emph{Nature communications} \textbf{2013}, \emph{4}, 2256\relax
\mciteBstWouldAddEndPuncttrue
\mciteSetBstMidEndSepPunct{\mcitedefaultmidpunct}
{\mcitedefaultendpunct}{\mcitedefaultseppunct}\relax
\EndOfBibitem
\bibitem[Naaman and Waldeck(2015)Naaman, and Waldeck]{naaman2015spintronics}
Naaman,~R.; Waldeck,~D.~H. Spintronics and chirality: Spin selectivity in electron transport through chiral molecules. \emph{Annual review of physical chemistry} \textbf{2015}, \emph{66}, 263--281\relax
\mciteBstWouldAddEndPuncttrue
\mciteSetBstMidEndSepPunct{\mcitedefaultmidpunct}
{\mcitedefaultendpunct}{\mcitedefaultseppunct}\relax
\EndOfBibitem
\bibitem[Liang \latin{et~al.}(2022)Liang, Banjac, Martin, Zigon, Lee, Vanthuyne, Garc{\'e}s-Pineda, Gal{\'a}n-Mascar{\'o}s, Hu, Avarvari, \latin{et~al.} others]{liang2022enhancement}
Liang,~Y.; Banjac,~K.; Martin,~K.; Zigon,~N.; Lee,~S.; Vanthuyne,~N.; Garc{\'e}s-Pineda,~F.~A.; Gal{\'a}n-Mascar{\'o}s,~J.~R.; Hu,~X.; Avarvari,~N.; others Enhancement of electrocatalytic oxygen evolution by chiral molecular functionalization of hybrid 2D electrodes. \emph{Nature Communications} \textbf{2022}, \emph{13}, 3356\relax
\mciteBstWouldAddEndPuncttrue
\mciteSetBstMidEndSepPunct{\mcitedefaultmidpunct}
{\mcitedefaultendpunct}{\mcitedefaultseppunct}\relax
\EndOfBibitem
\bibitem[Chang \latin{et~al.}(2018)Chang, Wieder, Schindler, Sanchez, Belopolski, Huang, Singh, Wu, Chang, Neupert, \latin{et~al.} others]{chang2018topological}
Chang,~G.; Wieder,~B.~J.; Schindler,~F.; Sanchez,~D.~S.; Belopolski,~I.; Huang,~S.-M.; Singh,~B.; Wu,~D.; Chang,~T.-R.; Neupert,~T.; others Topological quantum properties of chiral crystals. \emph{Nature materials} \textbf{2018}, \emph{17}, 978--985\relax
\mciteBstWouldAddEndPuncttrue
\mciteSetBstMidEndSepPunct{\mcitedefaultmidpunct}
{\mcitedefaultendpunct}{\mcitedefaultseppunct}\relax
\EndOfBibitem
\bibitem[Hasan \latin{et~al.}(2021)Hasan, Chang, Belopolski, Bian, Xu, and Yin]{hasan2021weyl}
Hasan,~M.~Z.; Chang,~G.; Belopolski,~I.; Bian,~G.; Xu,~S.-Y.; Yin,~J.-X. Weyl, Dirac and high-fold chiral fermions in topological quantum matter. \emph{Nature Reviews Materials} \textbf{2021}, \emph{6}, 784--803\relax
\mciteBstWouldAddEndPuncttrue
\mciteSetBstMidEndSepPunct{\mcitedefaultmidpunct}
{\mcitedefaultendpunct}{\mcitedefaultseppunct}\relax
\EndOfBibitem
\bibitem[Yang \latin{et~al.}(2023)Yang, Xiao, Robredo, Vergniory, Yan, and Felser]{yang2023monopole}
Yang,~Q.; Xiao,~J.; Robredo,~I.; Vergniory,~M.~G.; Yan,~B.; Felser,~C. Monopole-like orbital-momentum locking and the induced orbital transport in topological chiral semimetals. \emph{Proceedings of the National Academy of Sciences} \textbf{2023}, \emph{120}, e2305541120\relax
\mciteBstWouldAddEndPuncttrue
\mciteSetBstMidEndSepPunct{\mcitedefaultmidpunct}
{\mcitedefaultendpunct}{\mcitedefaultseppunct}\relax
\EndOfBibitem
\bibitem[Zhang and Niu(2015)Zhang, and Niu]{zhang2015chiral}
Zhang,~L.; Niu,~Q. Chiral phonons at high-symmetry points in monolayer hexagonal lattices. \emph{Physical review letters} \textbf{2015}, \emph{115}, 115502\relax
\mciteBstWouldAddEndPuncttrue
\mciteSetBstMidEndSepPunct{\mcitedefaultmidpunct}
{\mcitedefaultendpunct}{\mcitedefaultseppunct}\relax
\EndOfBibitem
\bibitem[Zhu \latin{et~al.}(2018)Zhu, Yi, Li, Xiao, Zhang, Yang, Kaindl, Li, Wang, and Zhang]{zhu2018observation}
Zhu,~H.; Yi,~J.; Li,~M.-Y.; Xiao,~J.; Zhang,~L.; Yang,~C.-W.; Kaindl,~R.~A.; Li,~L.-J.; Wang,~Y.; Zhang,~X. Observation of chiral phonons. \emph{Science} \textbf{2018}, \emph{359}, 579--582\relax
\mciteBstWouldAddEndPuncttrue
\mciteSetBstMidEndSepPunct{\mcitedefaultmidpunct}
{\mcitedefaultendpunct}{\mcitedefaultseppunct}\relax
\EndOfBibitem
\bibitem[Ishito \latin{et~al.}(2023)Ishito, Mao, Kousaka, Togawa, Iwasaki, Zhang, Murakami, Kishine, and Satoh]{ishito2023truly}
Ishito,~K.; Mao,~H.; Kousaka,~Y.; Togawa,~Y.; Iwasaki,~S.; Zhang,~T.; Murakami,~S.; Kishine,~J.-i.; Satoh,~T. Truly chiral phonons in $\alpha$-HgS. \emph{Nature Physics} \textbf{2023}, \emph{19}, 35--39\relax
\mciteBstWouldAddEndPuncttrue
\mciteSetBstMidEndSepPunct{\mcitedefaultmidpunct}
{\mcitedefaultendpunct}{\mcitedefaultseppunct}\relax
\EndOfBibitem
\bibitem[Ueda \latin{et~al.}(2023)Ueda, Garc{\'\i}a-Fern{\'a}ndez, Agrestini, Romao, van~den Brink, Spaldin, Zhou, and Staub]{ueda2023chiral}
Ueda,~H.; Garc{\'\i}a-Fern{\'a}ndez,~M.; Agrestini,~S.; Romao,~C.~P.; van~den Brink,~J.; Spaldin,~N.~A.; Zhou,~K.-J.; Staub,~U. Chiral phonons in quartz probed by X-rays. \emph{Nature} \textbf{2023}, 1--5\relax
\mciteBstWouldAddEndPuncttrue
\mciteSetBstMidEndSepPunct{\mcitedefaultmidpunct}
{\mcitedefaultendpunct}{\mcitedefaultseppunct}\relax
\EndOfBibitem
\bibitem[Romao and Juraschek(2023)Romao, and Juraschek]{romao2023phonon}
Romao,~C.~P.; Juraschek,~D.~M. Phonon-induced geometric chirality. \emph{arXiv preprint (cond-mat.mtrl-sci)} \textbf{2023}, \emph{arXiv:2311.00824}, url: https://arxiv.org/abs/2311.00824\relax
\mciteBstWouldAddEndPuncttrue
\mciteSetBstMidEndSepPunct{\mcitedefaultmidpunct}
{\mcitedefaultendpunct}{\mcitedefaultseppunct}\relax
\EndOfBibitem
\bibitem[Lange \latin{et~al.}(2023)Lange, Pottecher, Robey, Monserrat, and Peng]{lange2023negative}
Lange,~G.~F.; Pottecher,~J.~D.; Robey,~C.; Monserrat,~B.; Peng,~B. Negative refraction of Weyl phonons at twin quartz interfaces. \emph{ACS Materials Letters} \textbf{2023}, \emph{6}, 847--855\relax
\mciteBstWouldAddEndPuncttrue
\mciteSetBstMidEndSepPunct{\mcitedefaultmidpunct}
{\mcitedefaultendpunct}{\mcitedefaultseppunct}\relax
\EndOfBibitem
\bibitem[Hu \latin{et~al.}(2020)Hu, Florio, Chen, Phelan, Siegler, Zhou, Guo, Hawks, Jiang, Feng, \latin{et~al.} others]{hu2020chiral}
Hu,~Y.; Florio,~F.; Chen,~Z.; Phelan,~W.~A.; Siegler,~M.~A.; Zhou,~Z.; Guo,~Y.; Hawks,~R.; Jiang,~J.; Feng,~J.; others A chiral switchable photovoltaic ferroelectric 1D perovskite. \emph{Science advances} \textbf{2020}, \emph{6}, eaay4213\relax
\mciteBstWouldAddEndPuncttrue
\mciteSetBstMidEndSepPunct{\mcitedefaultmidpunct}
{\mcitedefaultendpunct}{\mcitedefaultseppunct}\relax
\EndOfBibitem
\bibitem[Fu \latin{et~al.}(2022)Fu, Hou, He, Liu, Lv, and Tang]{fu2022multiaxial}
Fu,~D.; Hou,~Z.; He,~Y.; Liu,~J.-C.; Lv,~H.-P.; Tang,~Y.-Y. Multiaxial ferroelectricity and ferroelasticity in a chiral perovskite. \emph{Chemistry of Materials} \textbf{2022}, \emph{34}, 3518--3524\relax
\mciteBstWouldAddEndPuncttrue
\mciteSetBstMidEndSepPunct{\mcitedefaultmidpunct}
{\mcitedefaultendpunct}{\mcitedefaultseppunct}\relax
\EndOfBibitem
\bibitem[Das \latin{et~al.}(2024)Das, Swain, Mahata, Prajapat, Upadhyay, Saikia, Reddy, De~Angelis, and Sarma]{das2024family}
Das,~R.; Swain,~D.; Mahata,~A.; Prajapat,~D.; Upadhyay,~S.~K.; Saikia,~S.; Reddy,~V.~R.; De~Angelis,~F.; Sarma,~D. Family of Chiral Ferroelectric Compounds with Widely Tunable Band Gaps. \emph{Chemistry of Materials} \textbf{2024}, \emph{36}, 1891--1898\relax
\mciteBstWouldAddEndPuncttrue
\mciteSetBstMidEndSepPunct{\mcitedefaultmidpunct}
{\mcitedefaultendpunct}{\mcitedefaultseppunct}\relax
\EndOfBibitem
\bibitem[Chien(2013)]{chien2013hall}
Chien,~C. \emph{The Hall effect and its applications}; Springer Science \& Business Media, 2013; ISBN: 978-1-4757-1369-5, 978-1-4757-1367-1\relax
\mciteBstWouldAddEndPuncttrue
\mciteSetBstMidEndSepPunct{\mcitedefaultmidpunct}
{\mcitedefaultendpunct}{\mcitedefaultseppunct}\relax
\EndOfBibitem
\bibitem[Cage \latin{et~al.}(2012)Cage, Klitzing, Chang, Duncan, Haldane, Laughlin, Pruisken, and Thouless]{cage2012quantum}
Cage,~M.~E.; Klitzing,~K.; Chang,~A.; Duncan,~F.; Haldane,~M.; Laughlin,~R.~B.; Pruisken,~A.; Thouless,~D. \emph{The quantum Hall effect}; Springer Science \& Business Media, 2012; ISBN: 146123350X, 9781461233503\relax
\mciteBstWouldAddEndPuncttrue
\mciteSetBstMidEndSepPunct{\mcitedefaultmidpunct}
{\mcitedefaultendpunct}{\mcitedefaultseppunct}\relax
\EndOfBibitem
\bibitem[Maciejko \latin{et~al.}(2011)Maciejko, Hughes, and Zhang]{maciejko2011quantum}
Maciejko,~J.; Hughes,~T.~L.; Zhang,~S.-C. The quantum spin Hall effect. \emph{Annu. Rev. Condens. Matter Phys.} \textbf{2011}, \emph{2}, 31--53\relax
\mciteBstWouldAddEndPuncttrue
\mciteSetBstMidEndSepPunct{\mcitedefaultmidpunct}
{\mcitedefaultendpunct}{\mcitedefaultseppunct}\relax
\EndOfBibitem
\bibitem[Chang \latin{et~al.}(2023)Chang, Liu, and MacDonald]{chang2023colloquium}
Chang,~C.-Z.; Liu,~C.-X.; MacDonald,~A.~H. Colloquium: Quantum anomalous hall effect. \emph{Reviews of Modern Physics} \textbf{2023}, \emph{95}, 011002\relax
\mciteBstWouldAddEndPuncttrue
\mciteSetBstMidEndSepPunct{\mcitedefaultmidpunct}
{\mcitedefaultendpunct}{\mcitedefaultseppunct}\relax
\EndOfBibitem
\bibitem[Du \latin{et~al.}(2021)Du, Lu, and Xie]{du2021nonlinear}
Du,~Z.; Lu,~H.-Z.; Xie,~X. Nonlinear hall effects. \emph{Nature Reviews Physics} \textbf{2021}, \emph{3}, 744--752\relax
\mciteBstWouldAddEndPuncttrue
\mciteSetBstMidEndSepPunct{\mcitedefaultmidpunct}
{\mcitedefaultendpunct}{\mcitedefaultseppunct}\relax
\EndOfBibitem
\bibitem[Ortix(2021)]{ortix2021nonlinear}
Ortix,~C. Nonlinear Hall Effect with Time-Reversal Symmetry: Theory and Material Realizations. \emph{Advanced Quantum Technologies} \textbf{2021}, \emph{4}, 2100056\relax
\mciteBstWouldAddEndPuncttrue
\mciteSetBstMidEndSepPunct{\mcitedefaultmidpunct}
{\mcitedefaultendpunct}{\mcitedefaultseppunct}\relax
\EndOfBibitem
\bibitem[Bandyopadhyay \latin{et~al.}(2024)Bandyopadhyay, Joseph, and Narayan]{bandyopadhyay2024non}
Bandyopadhyay,~A.; Joseph,~N.~B.; Narayan,~A. Non-linear Hall effects: Mechanisms and materials. \emph{Materials Today Electronics} \textbf{2024}, \emph{8}, 100101\relax
\mciteBstWouldAddEndPuncttrue
\mciteSetBstMidEndSepPunct{\mcitedefaultmidpunct}
{\mcitedefaultendpunct}{\mcitedefaultseppunct}\relax
\EndOfBibitem
\bibitem[Sodemann and Fu(2015)Sodemann, and Fu]{sodemann2015quantum}
Sodemann,~I.; Fu,~L. Quantum nonlinear Hall effect induced by Berry curvature dipole in time-reversal invariant materials. \emph{Physical review letters} \textbf{2015}, \emph{115}, 216806\relax
\mciteBstWouldAddEndPuncttrue
\mciteSetBstMidEndSepPunct{\mcitedefaultmidpunct}
{\mcitedefaultendpunct}{\mcitedefaultseppunct}\relax
\EndOfBibitem
\bibitem[Son \latin{et~al.}(2019)Son, Kim, Ahn, Lee, and Lee]{son2019strain}
Son,~J.; Kim,~K.-H.; Ahn,~Y.; Lee,~H.-W.; Lee,~J. Strain engineering of the Berry curvature dipole and valley magnetization in monolayer MoS 2. \emph{Physical review letters} \textbf{2019}, \emph{123}, 036806\relax
\mciteBstWouldAddEndPuncttrue
\mciteSetBstMidEndSepPunct{\mcitedefaultmidpunct}
{\mcitedefaultendpunct}{\mcitedefaultseppunct}\relax
\EndOfBibitem
\bibitem[You \latin{et~al.}(2018)You, Fang, Xu, Kaxiras, and Low]{you2018berry}
You,~J.-S.; Fang,~S.; Xu,~S.-Y.; Kaxiras,~E.; Low,~T. Berry curvature dipole current in the transition metal dichalcogenides family. \emph{Physical Review B} \textbf{2018}, \emph{98}, 121109\relax
\mciteBstWouldAddEndPuncttrue
\mciteSetBstMidEndSepPunct{\mcitedefaultmidpunct}
{\mcitedefaultendpunct}{\mcitedefaultseppunct}\relax
\EndOfBibitem
\bibitem[Xiao \latin{et~al.}(2020)Xiao, Shao, Zhang, and Jiang]{xiao2020two}
Xiao,~R.-C.; Shao,~D.-F.; Zhang,~Z.-Q.; Jiang,~H. Two-dimensional metals for piezoelectriclike devices based on Berry-curvature dipole. \emph{Physical Review Applied} \textbf{2020}, \emph{13}, 044014\relax
\mciteBstWouldAddEndPuncttrue
\mciteSetBstMidEndSepPunct{\mcitedefaultmidpunct}
{\mcitedefaultendpunct}{\mcitedefaultseppunct}\relax
\EndOfBibitem
\bibitem[Zhou \latin{et~al.}(2020)Zhou, Zhang, and Law]{zhou2020highly}
Zhou,~B.~T.; Zhang,~C.-P.; Law,~K.~T. Highly tunable nonlinear Hall effects induced by spin-orbit couplings in strained polar transition-metal dichalcogenides. \emph{Physical Review Applied} \textbf{2020}, \emph{13}, 024053\relax
\mciteBstWouldAddEndPuncttrue
\mciteSetBstMidEndSepPunct{\mcitedefaultmidpunct}
{\mcitedefaultendpunct}{\mcitedefaultseppunct}\relax
\EndOfBibitem
\bibitem[Joseph \latin{et~al.}(2021)Joseph, Roy, and Narayan]{joseph2021tunable}
Joseph,~N.~B.; Roy,~S.; Narayan,~A. Tunable topology and berry curvature dipole in transition metal dichalcogenide Janus monolayers. \emph{Materials Research Express} \textbf{2021}, \emph{8}, 124001\relax
\mciteBstWouldAddEndPuncttrue
\mciteSetBstMidEndSepPunct{\mcitedefaultmidpunct}
{\mcitedefaultendpunct}{\mcitedefaultseppunct}\relax
\EndOfBibitem
\bibitem[Zhang \latin{et~al.}(2018)Zhang, van~den Brink, Felser, and Yan]{zhang2018electrically}
Zhang,~Y.; van~den Brink,~J.; Felser,~C.; Yan,~B. Electrically tuneable nonlinear anomalous Hall effect in two-dimensional transition-metal dichalcogenides WTe2 and MoTe2. \emph{2D Materials} \textbf{2018}, \emph{5}, 044001\relax
\mciteBstWouldAddEndPuncttrue
\mciteSetBstMidEndSepPunct{\mcitedefaultmidpunct}
{\mcitedefaultendpunct}{\mcitedefaultseppunct}\relax
\EndOfBibitem
\bibitem[Joseph and Narayan(2021)Joseph, and Narayan]{joseph2021topological}
Joseph,~N.~B.; Narayan,~A. Topological properties of bulk and bilayer 2M WS2: a first-principles study. \emph{Journal of Physics: Condensed Matter} \textbf{2021}, \emph{33}, 465001\relax
\mciteBstWouldAddEndPuncttrue
\mciteSetBstMidEndSepPunct{\mcitedefaultmidpunct}
{\mcitedefaultendpunct}{\mcitedefaultseppunct}\relax
\EndOfBibitem
\bibitem[He and Weng(2021)He, and Weng]{he2021giant}
He,~Z.; Weng,~H. Giant nonlinear Hall effect in twisted bilayer WTe2. \emph{npj Quantum Materials} \textbf{2021}, \emph{6}, 101\relax
\mciteBstWouldAddEndPuncttrue
\mciteSetBstMidEndSepPunct{\mcitedefaultmidpunct}
{\mcitedefaultendpunct}{\mcitedefaultseppunct}\relax
\EndOfBibitem
\bibitem[Jin \latin{et~al.}(2021)Jin, Su, Li, Yu, Guo, and Wei]{jin2021strain}
Jin,~H.; Su,~H.; Li,~X.; Yu,~Y.; Guo,~H.; Wei,~Y. Strain-gated nonlinear Hall effect in two-dimensional MoSe 2/WSe 2 van der Waals heterostructure. \emph{Physical Review B} \textbf{2021}, \emph{104}, 195404\relax
\mciteBstWouldAddEndPuncttrue
\mciteSetBstMidEndSepPunct{\mcitedefaultmidpunct}
{\mcitedefaultendpunct}{\mcitedefaultseppunct}\relax
\EndOfBibitem
\bibitem[Zhang \latin{et~al.}(2018)Zhang, Sun, and Yan]{zhang2018berry}
Zhang,~Y.; Sun,~Y.; Yan,~B. Berry curvature dipole in Weyl semimetal materials: an ab initio study. \emph{Physical Review B} \textbf{2018}, \emph{97}, 041101\relax
\mciteBstWouldAddEndPuncttrue
\mciteSetBstMidEndSepPunct{\mcitedefaultmidpunct}
{\mcitedefaultendpunct}{\mcitedefaultseppunct}\relax
\EndOfBibitem
\bibitem[Zeng \latin{et~al.}(2021)Zeng, Nandy, and Tewari]{zeng2021nonlinear}
Zeng,~C.; Nandy,~S.; Tewari,~S. Nonlinear transport in Weyl semimetals induced by Berry curvature dipole. \emph{Physical Review B} \textbf{2021}, \emph{103}, 245119\relax
\mciteBstWouldAddEndPuncttrue
\mciteSetBstMidEndSepPunct{\mcitedefaultmidpunct}
{\mcitedefaultendpunct}{\mcitedefaultseppunct}\relax
\EndOfBibitem
\bibitem[Singh \latin{et~al.}(2020)Singh, Kim, Rabe, and Vanderbilt]{singh2020engineering}
Singh,~S.; Kim,~J.; Rabe,~K.~M.; Vanderbilt,~D. Engineering Weyl phases and nonlinear Hall effects in T d-MoTe 2. \emph{Physical review letters} \textbf{2020}, \emph{125}, 046402\relax
\mciteBstWouldAddEndPuncttrue
\mciteSetBstMidEndSepPunct{\mcitedefaultmidpunct}
{\mcitedefaultendpunct}{\mcitedefaultseppunct}\relax
\EndOfBibitem
\bibitem[Pang \latin{et~al.}(2024)Pang, Jin, and He]{pang2024tuning}
Pang,~H.; Jin,~G.; He,~L. Tuning of Berry-curvature dipole in TaAs slabs: An effective route to enhance the nonlinear Hall response. \emph{Physical Review Materials} \textbf{2024}, \emph{8}, 043403\relax
\mciteBstWouldAddEndPuncttrue
\mciteSetBstMidEndSepPunct{\mcitedefaultmidpunct}
{\mcitedefaultendpunct}{\mcitedefaultseppunct}\relax
\EndOfBibitem
\bibitem[Chen \latin{et~al.}(2019)Chen, Wang, Wang, and Zhang]{chen2019strain}
Chen,~C.; Wang,~H.; Wang,~D.; Zhang,~H. Strain-engineered nonlinear Hall effect in HgTe. Spin. 2019; p 1940017\relax
\mciteBstWouldAddEndPuncttrue
\mciteSetBstMidEndSepPunct{\mcitedefaultmidpunct}
{\mcitedefaultendpunct}{\mcitedefaultseppunct}\relax
\EndOfBibitem
\bibitem[Bandyopadhyay \latin{et~al.}(2022)Bandyopadhyay, Joseph, and Narayan]{bandyopadhyay2022electrically}
Bandyopadhyay,~A.; Joseph,~N.~B.; Narayan,~A. Electrically switchable giant Berry curvature dipole in silicene, germanene and stanene. \emph{2D Materials} \textbf{2022}, \emph{9}, 035013\relax
\mciteBstWouldAddEndPuncttrue
\mciteSetBstMidEndSepPunct{\mcitedefaultmidpunct}
{\mcitedefaultendpunct}{\mcitedefaultseppunct}\relax
\EndOfBibitem
\bibitem[Bandyopadhyay \latin{et~al.}(2023)Bandyopadhyay, Joseph, and Narayan]{bandyopadhyay2023berry}
Bandyopadhyay,~A.; Joseph,~N.~B.; Narayan,~A. Berry curvature dipole and its strain engineering in layered phosphorene. \emph{Materials Today Electronics} \textbf{2023}, \emph{6}, 100076\relax
\mciteBstWouldAddEndPuncttrue
\mciteSetBstMidEndSepPunct{\mcitedefaultmidpunct}
{\mcitedefaultendpunct}{\mcitedefaultseppunct}\relax
\EndOfBibitem
\bibitem[Kiswandhi and Osada(2021)Kiswandhi, and Osada]{kiswandhi2021observation}
Kiswandhi,~A.; Osada,~T. Observation of possible nonlinear anomalous Hall effect in organic two-dimensional Dirac fermion system. \emph{Journal of Physics: Condensed Matter} \textbf{2021}, \emph{34}, 105602\relax
\mciteBstWouldAddEndPuncttrue
\mciteSetBstMidEndSepPunct{\mcitedefaultmidpunct}
{\mcitedefaultendpunct}{\mcitedefaultseppunct}\relax
\EndOfBibitem
\bibitem[Ma \latin{et~al.}(2019)Ma, Xu, Shen, MacNeill, Fatemi, Chang, Mier~Valdivia, Wu, Du, Hsu, \latin{et~al.} others]{ma2019observation}
Ma,~Q.; Xu,~S.-Y.; Shen,~H.; MacNeill,~D.; Fatemi,~V.; Chang,~T.-R.; Mier~Valdivia,~A.~M.; Wu,~S.; Du,~Z.; Hsu,~C.-H.; others Observation of the nonlinear Hall effect under time-reversal-symmetric conditions. \emph{Nature} \textbf{2019}, \emph{565}, 337--342\relax
\mciteBstWouldAddEndPuncttrue
\mciteSetBstMidEndSepPunct{\mcitedefaultmidpunct}
{\mcitedefaultendpunct}{\mcitedefaultseppunct}\relax
\EndOfBibitem
\bibitem[Kang \latin{et~al.}(2019)Kang, Li, Sohn, Shan, and Mak]{kang2019nonlinear}
Kang,~K.; Li,~T.; Sohn,~E.; Shan,~J.; Mak,~K.~F. Nonlinear anomalous Hall effect in few-layer WTe2. \emph{Nature materials} \textbf{2019}, \emph{18}, 324--328\relax
\mciteBstWouldAddEndPuncttrue
\mciteSetBstMidEndSepPunct{\mcitedefaultmidpunct}
{\mcitedefaultendpunct}{\mcitedefaultseppunct}\relax
\EndOfBibitem
\bibitem[Ma \latin{et~al.}(2022)Ma, Chen, Yananose, Zhou, Wang, Li, Zhu, Wu, Xu, Yu, \latin{et~al.} others]{ma2022growth}
Ma,~T.; Chen,~H.; Yananose,~K.; Zhou,~X.; Wang,~L.; Li,~R.; Zhu,~Z.; Wu,~Z.; Xu,~Q.-H.; Yu,~J.; others Growth of bilayer MoTe2 single crystals with strong non-linear Hall effect. \emph{Nature Communications} \textbf{2022}, \emph{13}, 5465\relax
\mciteBstWouldAddEndPuncttrue
\mciteSetBstMidEndSepPunct{\mcitedefaultmidpunct}
{\mcitedefaultendpunct}{\mcitedefaultseppunct}\relax
\EndOfBibitem
\bibitem[Huang \latin{et~al.}(2023)Huang, Wu, Hu, Cai, Li, An, Feng, Ye, Lin, Law, \latin{et~al.} others]{huang2023giant}
Huang,~M.; Wu,~Z.; Hu,~J.; Cai,~X.; Li,~E.; An,~L.; Feng,~X.; Ye,~Z.; Lin,~N.; Law,~K.~T.; others Giant nonlinear Hall effect in twisted bilayer WSe2. \emph{National Science Review} \textbf{2023}, \emph{10}, nwac232\relax
\mciteBstWouldAddEndPuncttrue
\mciteSetBstMidEndSepPunct{\mcitedefaultmidpunct}
{\mcitedefaultendpunct}{\mcitedefaultseppunct}\relax
\EndOfBibitem
\bibitem[Kang \latin{et~al.}(2023)Kang, Zhao, Zeng, Watanabe, Taniguchi, Shan, and Mak]{kang2023switchable}
Kang,~K.; Zhao,~W.; Zeng,~Y.; Watanabe,~K.; Taniguchi,~T.; Shan,~J.; Mak,~K.~F. Switchable moir{\'e} potentials in ferroelectric WTe2/WSe2 superlattices. \emph{Nature Nanotechnology} \textbf{2023}, 1--6\relax
\mciteBstWouldAddEndPuncttrue
\mciteSetBstMidEndSepPunct{\mcitedefaultmidpunct}
{\mcitedefaultendpunct}{\mcitedefaultseppunct}\relax
\EndOfBibitem
\bibitem[Ho \latin{et~al.}(2021)Ho, Chang, Hsieh, Lo, Huang, Vu, Ortix, and Chen]{ho2021hall}
Ho,~S.-C.; Chang,~C.-H.; Hsieh,~Y.-C.; Lo,~S.-T.; Huang,~B.; Vu,~T.-H.-Y.; Ortix,~C.; Chen,~T.-M. Hall effects in artificially corrugated bilayer graphene without breaking time-reversal symmetry. \emph{Nature Electronics} \textbf{2021}, \emph{4}, 116--125\relax
\mciteBstWouldAddEndPuncttrue
\mciteSetBstMidEndSepPunct{\mcitedefaultmidpunct}
{\mcitedefaultendpunct}{\mcitedefaultseppunct}\relax
\EndOfBibitem
\bibitem[Huang \latin{et~al.}(2023)Huang, Wu, Zhang, Feng, Zhou, Wang, Chen, Cheng, Sun, Meng, \latin{et~al.} others]{huang2023intrinsic}
Huang,~M.; Wu,~Z.; Zhang,~X.; Feng,~X.; Zhou,~Z.; Wang,~S.; Chen,~Y.; Cheng,~C.; Sun,~K.; Meng,~Z.~Y.; others Intrinsic nonlinear Hall effect and gate-switchable Berry curvature sliding in twisted bilayer graphene. \emph{Physical Review Letters} \textbf{2023}, \emph{131}, 066301\relax
\mciteBstWouldAddEndPuncttrue
\mciteSetBstMidEndSepPunct{\mcitedefaultmidpunct}
{\mcitedefaultendpunct}{\mcitedefaultseppunct}\relax
\EndOfBibitem
\bibitem[He \latin{et~al.}(2022)He, Koon, Isobe, Tan, Hu, Neto, Fu, and Yang]{he2022graphene}
He,~P.; Koon,~G. K.~W.; Isobe,~H.; Tan,~J.~Y.; Hu,~J.; Neto,~A. H.~C.; Fu,~L.; Yang,~H. Graphene moir{\'e} superlattices with giant quantum nonlinearity of chiral Bloch electrons. \emph{Nature nanotechnology} \textbf{2022}, \emph{17}, 378--383\relax
\mciteBstWouldAddEndPuncttrue
\mciteSetBstMidEndSepPunct{\mcitedefaultmidpunct}
{\mcitedefaultendpunct}{\mcitedefaultseppunct}\relax
\EndOfBibitem
\bibitem[Sinha \latin{et~al.}(2022)Sinha, Adak, Chakraborty, Das, Debnath, Sangani, Watanabe, Taniguchi, Waghmare, Agarwal, \latin{et~al.} others]{sinha2022berry}
Sinha,~S.; Adak,~P.~C.; Chakraborty,~A.; Das,~K.; Debnath,~K.; Sangani,~L.~V.; Watanabe,~K.; Taniguchi,~T.; Waghmare,~U.~V.; Agarwal,~A.; others Berry curvature dipole senses topological transition in a moir{\'e} superlattice. \emph{Nature Physics} \textbf{2022}, \emph{18}, 765--770\relax
\mciteBstWouldAddEndPuncttrue
\mciteSetBstMidEndSepPunct{\mcitedefaultmidpunct}
{\mcitedefaultendpunct}{\mcitedefaultseppunct}\relax
\EndOfBibitem
\bibitem[Zhong \latin{et~al.}(2023)Zhong, Duan, Zhang, Peng, Feng, Hu, Wang, Mao, Liu, and Yao]{zhong2023effective}
Zhong,~J.; Duan,~J.; Zhang,~S.; Peng,~H.; Feng,~Q.; Hu,~Y.; Wang,~Q.; Mao,~J.; Liu,~J.; Yao,~Y. Effective manipulation and realization of a colossal nonlinear Hall effect in an electric-field tunable moir$\backslash$'e system. \emph{arXiv preprint (cond-mat.mes-hall)} \textbf{2023}, \emph{arXiv:2301.12117}, url: https://arxiv.org/abs/2301.12117\relax
\mciteBstWouldAddEndPuncttrue
\mciteSetBstMidEndSepPunct{\mcitedefaultmidpunct}
{\mcitedefaultendpunct}{\mcitedefaultseppunct}\relax
\EndOfBibitem
\bibitem[Zhang \latin{et~al.}(2022)Zhang, Liang, Kaneko, Nagaosa, and Tokura]{zhang2022pst}
Zhang,~C.-L.; Liang,~T.; Kaneko,~Y.; Nagaosa,~N.; Tokura,~Y. Giant Berry curvature dipole density in a ferroelectric Weyl semimetal. \emph{npj Quantum Materials} \textbf{2022}, \emph{7}, 103\relax
\mciteBstWouldAddEndPuncttrue
\mciteSetBstMidEndSepPunct{\mcitedefaultmidpunct}
{\mcitedefaultendpunct}{\mcitedefaultseppunct}\relax
\EndOfBibitem
\bibitem[Nishijima \latin{et~al.}(2023)Nishijima, Watanabe, Sekiguchi, Ando, Shigematsu, Ohshima, Kuroda, and Shiraishi]{nishijima2023ferroic}
Nishijima,~T.; Watanabe,~T.; Sekiguchi,~H.; Ando,~Y.; Shigematsu,~E.; Ohshima,~R.; Kuroda,~S.; Shiraishi,~M. Ferroic Berry Curvature Dipole in a Topological Crystalline Insulator at Room Temperature. \emph{Nano Letters} \textbf{2023}, \emph{23}, 2247--2252\relax
\mciteBstWouldAddEndPuncttrue
\mciteSetBstMidEndSepPunct{\mcitedefaultmidpunct}
{\mcitedefaultendpunct}{\mcitedefaultseppunct}\relax
\EndOfBibitem
\bibitem[Shvetsov \latin{et~al.}(2019)Shvetsov, Esin, Timonina, Kolesnikov, and Deviatov]{shvetsov2019nonlinear}
Shvetsov,~O.~O.; Esin,~V.~D.; Timonina,~A.~V.; Kolesnikov,~N.~N.; Deviatov,~E. Nonlinear Hall effect in three-dimensional Weyl and Dirac semimetals. \emph{JETP Letters} \textbf{2019}, \emph{109}, 715--721\relax
\mciteBstWouldAddEndPuncttrue
\mciteSetBstMidEndSepPunct{\mcitedefaultmidpunct}
{\mcitedefaultendpunct}{\mcitedefaultseppunct}\relax
\EndOfBibitem
\bibitem[Kumar \latin{et~al.}(2021)Kumar, Hsu, Sharma, Chang, Yu, Wang, Eda, Liang, and Yang]{kumar2021room}
Kumar,~D.; Hsu,~C.-H.; Sharma,~R.; Chang,~T.-R.; Yu,~P.; Wang,~J.; Eda,~G.; Liang,~G.; Yang,~H. Room-temperature nonlinear Hall effect and wireless radiofrequency rectification in Weyl semimetal TaIrTe4. \emph{Nature Nanotechnology} \textbf{2021}, \emph{16}, 421--425\relax
\mciteBstWouldAddEndPuncttrue
\mciteSetBstMidEndSepPunct{\mcitedefaultmidpunct}
{\mcitedefaultendpunct}{\mcitedefaultseppunct}\relax
\EndOfBibitem
\bibitem[Zhang and Fu(2021)Zhang, and Fu]{zhang2021terahertz}
Zhang,~Y.; Fu,~L. Terahertz detection based on nonlinear Hall effect without magnetic field. \emph{Proceedings of the National Academy of Sciences} \textbf{2021}, \emph{118}, e2100736118\relax
\mciteBstWouldAddEndPuncttrue
\mciteSetBstMidEndSepPunct{\mcitedefaultmidpunct}
{\mcitedefaultendpunct}{\mcitedefaultseppunct}\relax
\EndOfBibitem
\bibitem[Ramsden(2011)]{ramsden2011hall}
Ramsden,~E. \emph{Hall-effect sensors: theory and application}; Elsevier, 2011; ISBN: 978-0-7506-7934-3\relax
\mciteBstWouldAddEndPuncttrue
\mciteSetBstMidEndSepPunct{\mcitedefaultmidpunct}
{\mcitedefaultendpunct}{\mcitedefaultseppunct}\relax
\EndOfBibitem
\bibitem[Giannozzi \latin{et~al.}(2017)Giannozzi, Andreussi, Brumme, Bunau, Nardelli, Calandra, Car, Cavazzoni, Ceresoli, Cococcioni, Colonna, Carnimeo, Corso, de~Gironcoli, Delugas, Jr, Ferretti, Floris, Fratesi, Fugallo, Gebauer, Gerstmann, Giustino, Gorni, Jia, Kawamura, Ko, Kokalj, Küçükbenli, Lazzeri, Marsili, Marzari, Mauri, Nguyen, Nguyen, de-la Roza, Paulatto, Poncé, Rocca, Sabatini, Santra, Schlipf, Seitsonen, Smogunov, Timrov, Thonhauser, Umari, Vast, Wu, and Baroni]{QE-2017}
Giannozzi,~P. \latin{et~al.}  Advanced capabilities for materials modelling with QUANTUM ESPRESSO. \emph{Journal of Physics: Condensed Matter} \textbf{2017}, \emph{29}, 465901\relax
\mciteBstWouldAddEndPuncttrue
\mciteSetBstMidEndSepPunct{\mcitedefaultmidpunct}
{\mcitedefaultendpunct}{\mcitedefaultseppunct}\relax
\EndOfBibitem
\bibitem[Giannozzi \latin{et~al.}(2009)Giannozzi, Baroni, Bonini, Calandra, Car, Cavazzoni, Ceresoli, Chiarotti, Cococcioni, Dabo, {Dal Corso}, de~Gironcoli, Fabris, Fratesi, Gebauer, Gerstmann, Gougoussis, Kokalj, Lazzeri, Martin-Samos, Marzari, Mauri, Mazzarello, Paolini, Pasquarello, Paulatto, Sbraccia, Scandolo, Sclauzero, Seitsonen, Smogunov, Umari, and Wentzcovitch]{QE-2009}
Giannozzi,~P. \latin{et~al.}  QUANTUM ESPRESSO: a modular and open-source software project for quantum simulations of materials. \emph{Journal of Physics: Condensed Matter} \textbf{2009}, \emph{21}, 395502 (19pp)\relax
\mciteBstWouldAddEndPuncttrue
\mciteSetBstMidEndSepPunct{\mcitedefaultmidpunct}
{\mcitedefaultendpunct}{\mcitedefaultseppunct}\relax
\EndOfBibitem
\bibitem[Vanderbilt(1990)]{PhysRevB.41.7892}
Vanderbilt,~D. Soft self-consistent pseudopotentials in a generalized eigenvalue formalism. \emph{Phys. Rev. B} \textbf{1990}, \emph{41}, 7892--7895\relax
\mciteBstWouldAddEndPuncttrue
\mciteSetBstMidEndSepPunct{\mcitedefaultmidpunct}
{\mcitedefaultendpunct}{\mcitedefaultseppunct}\relax
\EndOfBibitem
\bibitem[Perdew \latin{et~al.}(1996)Perdew, Burke, and Ernzerhof]{perdew1996generalized}
Perdew,~J.~P.; Burke,~K.; Ernzerhof,~M. Generalized gradient approximation made simple. \emph{Physical review letters} \textbf{1996}, \emph{77}, 3865\relax
\mciteBstWouldAddEndPuncttrue
\mciteSetBstMidEndSepPunct{\mcitedefaultmidpunct}
{\mcitedefaultendpunct}{\mcitedefaultseppunct}\relax
\EndOfBibitem
\bibitem[Heyd \latin{et~al.}(2003)Heyd, Scuseria, and Ernzerhof]{heyd2003hybrid}
Heyd,~J.; Scuseria,~G.~E.; Ernzerhof,~M. Hybrid functionals based on a screened Coulomb potential. \emph{The Journal of chemical physics} \textbf{2003}, \emph{118}, 8207--8215\relax
\mciteBstWouldAddEndPuncttrue
\mciteSetBstMidEndSepPunct{\mcitedefaultmidpunct}
{\mcitedefaultendpunct}{\mcitedefaultseppunct}\relax
\EndOfBibitem
\bibitem[Grimme(2006)]{DFT-D2}
Grimme,~S. Semiempirical GGA-type density functional constructed with a long-range dispersion correction. \emph{Journal of Computational Chemistry} \textbf{2006}, \emph{27}, 1787--1799\relax
\mciteBstWouldAddEndPuncttrue
\mciteSetBstMidEndSepPunct{\mcitedefaultmidpunct}
{\mcitedefaultendpunct}{\mcitedefaultseppunct}\relax
\EndOfBibitem
\bibitem[Mostofi \latin{et~al.}(2014)Mostofi, Yates, Pizzi, Lee, Souza, Vanderbilt, and Marzari]{mostofi2014updated}
Mostofi,~A.~A.; Yates,~J.~R.; Pizzi,~G.; Lee,~Y.-S.; Souza,~I.; Vanderbilt,~D.; Marzari,~N. An updated version of wannier90: A tool for obtaining maximally-localised Wannier functions. \emph{Computer Physics Communications} \textbf{2014}, \emph{185}, 2309--2310\relax
\mciteBstWouldAddEndPuncttrue
\mciteSetBstMidEndSepPunct{\mcitedefaultmidpunct}
{\mcitedefaultendpunct}{\mcitedefaultseppunct}\relax
\EndOfBibitem
\bibitem[Marzari and Vanderbilt(1997)Marzari, and Vanderbilt]{marzari1997maximally}
Marzari,~N.; Vanderbilt,~D. Maximally localized generalized Wannier functions for composite energy bands. \emph{Physical review B} \textbf{1997}, \emph{56}, 12847\relax
\mciteBstWouldAddEndPuncttrue
\mciteSetBstMidEndSepPunct{\mcitedefaultmidpunct}
{\mcitedefaultendpunct}{\mcitedefaultseppunct}\relax
\EndOfBibitem
\bibitem[Wang \latin{et~al.}(2014)Wang, Lazar, Park, Millis, and Marianetti]{wang2014selectively}
Wang,~R.; Lazar,~E.~A.; Park,~H.; Millis,~A.~J.; Marianetti,~C.~A. Selectively localized Wannier functions. \emph{Physical Review B} \textbf{2014}, \emph{90}, 165125\relax
\mciteBstWouldAddEndPuncttrue
\mciteSetBstMidEndSepPunct{\mcitedefaultmidpunct}
{\mcitedefaultendpunct}{\mcitedefaultseppunct}\relax
\EndOfBibitem
\bibitem[Tsirkin(2021)]{tsirkin2021high}
Tsirkin,~S.~S. High performance Wannier interpolation of Berry curvature and related quantities with WannierBerri code. \emph{npj Computational Materials} \textbf{2021}, \emph{7}, 33\relax
\mciteBstWouldAddEndPuncttrue
\mciteSetBstMidEndSepPunct{\mcitedefaultmidpunct}
{\mcitedefaultendpunct}{\mcitedefaultseppunct}\relax
\EndOfBibitem
\bibitem[Wu \latin{et~al.}(2018)Wu, Zhang, Song, Troyer, and Soluyanov]{WU2017}
Wu,~Q.; Zhang,~S.; Song,~H.-F.; Troyer,~M.; Soluyanov,~A.~A. WannierTools : An open-source software package for novel topological materials. \emph{Computer Physics Communications} \textbf{2018}, \emph{224}, 405 -- 416\relax
\mciteBstWouldAddEndPuncttrue
\mciteSetBstMidEndSepPunct{\mcitedefaultmidpunct}
{\mcitedefaultendpunct}{\mcitedefaultseppunct}\relax
\EndOfBibitem
\bibitem[Xiao \latin{et~al.}(2020)Xiao, Shao, Huang, and Jiang]{xiao2020electrical}
Xiao,~R.-C.; Shao,~D.-F.; Huang,~W.; Jiang,~H. Electrical detection of ferroelectriclike metals through the nonlinear Hall effect. \emph{Physical Review B} \textbf{2020}, \emph{102}, 024109\relax
\mciteBstWouldAddEndPuncttrue
\mciteSetBstMidEndSepPunct{\mcitedefaultmidpunct}
{\mcitedefaultendpunct}{\mcitedefaultseppunct}\relax
\EndOfBibitem
\bibitem[Tsirkin \latin{et~al.}(2018)Tsirkin, Puente, and Souza]{tsirkin2018gyrotropic}
Tsirkin,~S.~S.; Puente,~P.~A.; Souza,~I. Gyrotropic effects in trigonal tellurium studied from first principles. \emph{Physical Review B} \textbf{2018}, \emph{97}, 035158\relax
\mciteBstWouldAddEndPuncttrue
\mciteSetBstMidEndSepPunct{\mcitedefaultmidpunct}
{\mcitedefaultendpunct}{\mcitedefaultseppunct}\relax
\EndOfBibitem
\bibitem[Sakano \latin{et~al.}(2020)Sakano, Hirayama, Takahashi, Akebi, Nakayama, Kuroda, Taguchi, Yoshikawa, Miyamoto, Okuda, \latin{et~al.} others]{sakano2020radial}
Sakano,~M.; Hirayama,~M.; Takahashi,~T.; Akebi,~S.; Nakayama,~M.; Kuroda,~K.; Taguchi,~K.; Yoshikawa,~T.; Miyamoto,~K.; Okuda,~T.; others Radial spin texture in elemental tellurium with chiral crystal structure. \emph{Physical review letters} \textbf{2020}, \emph{124}, 136404\relax
\mciteBstWouldAddEndPuncttrue
\mciteSetBstMidEndSepPunct{\mcitedefaultmidpunct}
{\mcitedefaultendpunct}{\mcitedefaultseppunct}\relax
\EndOfBibitem
\bibitem[Liu \latin{et~al.}(2023)Liu, Souza, and Tsirkin]{liu2023electricalmagnetochiralanisotropytrigonal}
Liu,~X.; Souza,~I.; Tsirkin,~S.~S. Electrical magnetochiral anisotropy in trigonal tellurium from first principles. \emph{arXiv preprint (cond-mat.mtrl-sci)} \textbf{2023}, \emph{arXiv:2303.10164}, submitted 17 Mar 2023. url: https://arxiv.org/abs/2303.10164\relax
\mciteBstWouldAddEndPuncttrue
\mciteSetBstMidEndSepPunct{\mcitedefaultmidpunct}
{\mcitedefaultendpunct}{\mcitedefaultseppunct}\relax
\EndOfBibitem
\bibitem[Anzin \latin{et~al.}(1977)Anzin, Eremets, Kosichkin, Nadezhdinskii, and Shirokov]{anzin1977measurement}
Anzin,~V.; Eremets,~M.; Kosichkin,~Y.~V.; Nadezhdinskii,~A.; Shirokov,~A. Measurement of the energy gap in tellurium under pressure. \emph{physica status solidi (a)} \textbf{1977}, \emph{42}, 385--390\relax
\mciteBstWouldAddEndPuncttrue
\mciteSetBstMidEndSepPunct{\mcitedefaultmidpunct}
{\mcitedefaultendpunct}{\mcitedefaultseppunct}\relax
\EndOfBibitem
\bibitem[Hirayama \latin{et~al.}(2015)Hirayama, Okugawa, Ishibashi, Murakami, and Miyake]{hirayama2015weyl}
Hirayama,~M.; Okugawa,~R.; Ishibashi,~S.; Murakami,~S.; Miyake,~T. Weyl node and spin texture in trigonal tellurium and selenium. \emph{Physical review letters} \textbf{2015}, \emph{114}, 206401\relax
\mciteBstWouldAddEndPuncttrue
\mciteSetBstMidEndSepPunct{\mcitedefaultmidpunct}
{\mcitedefaultendpunct}{\mcitedefaultseppunct}\relax
\EndOfBibitem
\bibitem[Nakayama \latin{et~al.}(2017)Nakayama, Kuno, Yamauchi, Souma, Sugawara, Oguchi, Sato, and Takahashi]{nakayama2017band}
Nakayama,~K.; Kuno,~M.; Yamauchi,~K.; Souma,~S.; Sugawara,~K.; Oguchi,~T.; Sato,~T.; Takahashi,~T. Band splitting and Weyl nodes in trigonal tellurium studied by angle-resolved photoemission spectroscopy and density functional theory. \emph{Physical review B} \textbf{2017}, \emph{95}, 125204\relax
\mciteBstWouldAddEndPuncttrue
\mciteSetBstMidEndSepPunct{\mcitedefaultmidpunct}
{\mcitedefaultendpunct}{\mcitedefaultseppunct}\relax
\EndOfBibitem
\bibitem[Niu \latin{et~al.}(2023)Niu, Huang, Ghosh, Tan, Wang, Wu, Xu, and Ye]{niu2023tunable}
Niu,~C.; Huang,~S.; Ghosh,~N.; Tan,~P.; Wang,~M.; Wu,~W.; Xu,~X.; Ye,~P.~D. Tunable Circular Photogalvanic and Photovoltaic Effect in 2D Tellurium with Different Chirality. \emph{Nano Letters} \textbf{2023}, \emph{23}, 3599--3606\relax
\mciteBstWouldAddEndPuncttrue
\mciteSetBstMidEndSepPunct{\mcitedefaultmidpunct}
{\mcitedefaultendpunct}{\mcitedefaultseppunct}\relax
\EndOfBibitem
\bibitem[Su{\'a}rez-Rodr{\'\i}guez \latin{et~al.}(2024)Su{\'a}rez-Rodr{\'\i}guez, Mart{\'\i}n-Garc{\'\i}a, Skowro{\'n}ski, Calavalle, Tsirkin, Souza, De~Juan, Chuvilin, Fert, Gobbi, \latin{et~al.} others]{suarez2024odd}
Su{\'a}rez-Rodr{\'\i}guez,~M.; Mart{\'\i}n-Garc{\'\i}a,~B.; Skowro{\'n}ski,~W.; Calavalle,~F.; Tsirkin,~S.~S.; Souza,~I.; De~Juan,~F.; Chuvilin,~A.; Fert,~A.; Gobbi,~M.; others Odd Nonlinear Conductivity under Spatial Inversion in Chiral Tellurium. \emph{Physical Review Letters} \textbf{2024}, \emph{132}, 046303\relax
\mciteBstWouldAddEndPuncttrue
\mciteSetBstMidEndSepPunct{\mcitedefaultmidpunct}
{\mcitedefaultendpunct}{\mcitedefaultseppunct}\relax
\EndOfBibitem
\bibitem[Masse \latin{et~al.}(1978)Masse, Aicardi, and Leyris]{masse1978study}
Masse,~G.; Aicardi,~J.; Leyris,~J. Study of the yellow emission of natural $\alpha$-HgS. \emph{Journal of Luminescence} \textbf{1978}, \emph{17}, 29--48\relax
\mciteBstWouldAddEndPuncttrue
\mciteSetBstMidEndSepPunct{\mcitedefaultmidpunct}
{\mcitedefaultendpunct}{\mcitedefaultseppunct}\relax
\EndOfBibitem
\bibitem[Jeong \latin{et~al.}(2014)Jeong, Deng, Keuleyan, Liu, and Guyot-Sionnest]{jeong2014air}
Jeong,~K.~S.; Deng,~Z.; Keuleyan,~S.; Liu,~H.; Guyot-Sionnest,~P. Air-stable n-doped colloidal HgS quantum dots. \emph{The Journal of Physical Chemistry Letters} \textbf{2014}, \emph{5}, 1139--1143\relax
\mciteBstWouldAddEndPuncttrue
\mciteSetBstMidEndSepPunct{\mcitedefaultmidpunct}
{\mcitedefaultendpunct}{\mcitedefaultseppunct}\relax
\EndOfBibitem
\bibitem[Tang \latin{et~al.}(2017)Tang, Zhou, and Zhang]{tang2017multiple}
Tang,~P.; Zhou,~Q.; Zhang,~S.-C. Multiple types of topological fermions in transition metal silicides. \emph{Physical review letters} \textbf{2017}, \emph{119}, 206402\relax
\mciteBstWouldAddEndPuncttrue
\mciteSetBstMidEndSepPunct{\mcitedefaultmidpunct}
{\mcitedefaultendpunct}{\mcitedefaultseppunct}\relax
\EndOfBibitem
\bibitem[Rao \latin{et~al.}(2019)Rao, Li, Zhang, Tian, Li, Fu, Tang, Wang, Li, Fan, \latin{et~al.} others]{rao2019observation}
Rao,~Z.; Li,~H.; Zhang,~T.; Tian,~S.; Li,~C.; Fu,~B.; Tang,~C.; Wang,~L.; Li,~Z.; Fan,~W.; others Observation of unconventional chiral fermions with long Fermi arcs in CoSi. \emph{Nature} \textbf{2019}, \emph{567}, 496--499\relax
\mciteBstWouldAddEndPuncttrue
\mciteSetBstMidEndSepPunct{\mcitedefaultmidpunct}
{\mcitedefaultendpunct}{\mcitedefaultseppunct}\relax
\EndOfBibitem
\bibitem[Bose and Narayan(2021)Bose, and Narayan]{bose2021strain}
Bose,~A.; Narayan,~A. Strain-induced topological charge control in multifold fermion systems. \emph{Journal of Physics: Condensed Matter} \textbf{2021}, \emph{33}, 375002\relax
\mciteBstWouldAddEndPuncttrue
\mciteSetBstMidEndSepPunct{\mcitedefaultmidpunct}
{\mcitedefaultendpunct}{\mcitedefaultseppunct}\relax
\EndOfBibitem
\bibitem[Roy and Narayan(2022)Roy, and Narayan]{roy2022non}
Roy,~S.; Narayan,~A. Non-linear Hall effect in multi-Weyl semimetals. \emph{Journal of Physics: Condensed Matter} \textbf{2022}, \emph{34}, 385301\relax
\mciteBstWouldAddEndPuncttrue
\mciteSetBstMidEndSepPunct{\mcitedefaultmidpunct}
{\mcitedefaultendpunct}{\mcitedefaultseppunct}\relax
\EndOfBibitem
\bibitem[Saha and Narayan(2023)Saha, and Narayan]{saha2023nonlinear}
Saha,~S.; Narayan,~A. Nonlinear Hall effect in Rashba systems with hexagonal warping. \emph{Journal of Physics: Condensed Matter} \textbf{2023}, \emph{35}, 485301\relax
\mciteBstWouldAddEndPuncttrue
\mciteSetBstMidEndSepPunct{\mcitedefaultmidpunct}
{\mcitedefaultendpunct}{\mcitedefaultseppunct}\relax
\EndOfBibitem
\bibitem[Facio \latin{et~al.}(2018)Facio, Efremov, Koepernik, You, Sodemann, and Van Den~Brink]{facio2018strongly}
Facio,~J.~I.; Efremov,~D.; Koepernik,~K.; You,~J.-S.; Sodemann,~I.; Van Den~Brink,~J. Strongly enhanced Berry dipole at topological phase transitions in BiTeI. \emph{Physical review letters} \textbf{2018}, \emph{121}, 246403\relax
\mciteBstWouldAddEndPuncttrue
\mciteSetBstMidEndSepPunct{\mcitedefaultmidpunct}
{\mcitedefaultendpunct}{\mcitedefaultseppunct}\relax
\EndOfBibitem
\end{mcitethebibliography}

\end{document}